\documentclass[preprintnumbers,amsmath,amssymb,usenatbib,nofootinbib,superscriptaddress,showkeys,showpacs,10pt]{revtex4-2}

\usepackage{amsmath}
\usepackage{amsfonts,color}
\usepackage{amssymb,float}
\usepackage{graphicx}
\usepackage{subfigure}
\usepackage{appendix}
\usepackage[colorlinks]{hyperref}
\usepackage{xcolor}
\usepackage{orcidlink}
\usepackage{epsf}
\usepackage{bm}
\usepackage{epstopdf}
\usepackage{natbib}
\usepackage{lipsum}
\usepackage{multirow}
\usepackage{nicematrix}

\begin{document}

\title{Cosmological Evolution, Analytical Parametrization, and Dynamical Stability of a Bouncing Universe in Non-Minimal Kinetic Coupling Gravity}

\author{Alireza Amani\orcidlink{0000-0002-1296-614X}}
\email{ al.amani@iau.ac.ir}
\affiliation{Department of Physics, Am.C., Islamic Azad University, Amol, Iran.}

\author{A. S. Kubeka\orcidlink{0000-0002-9185-7549}}
\email{ kubekas@unisa.ac.za}
\affiliation{Department of Mathematical Sciences, University of South Africa, Science Campus, P.O.Box 392, Florida, South Africa.}

\author{E. Mahichi\orcidlink{0000-0001-7737-0040}}
\email{elhammahichi@gmail.com}
\affiliation{Department of Physics, Am.C., Islamic Azad University, Amol, Iran.}

\date{\today}

\begin{abstract}

In this paper, the bouncing cosmology is studied in the framework of the non-minimal kinetic coupling theory between the scalar field and the Einstein tensor. After deriving the modified Friedmann equations and the field equation, by applying the Bounce condition, the system of equations is solved numerically without assuming any analytical form for the scale factor. Thus, the time evolution of the scale factor, the scalar field, and other cosmological quantities are obtained directly from the model dynamics. Then, in order to provide an explicit form for the cosmic evolution, an analytical parameterization for the scale factor is introduced, which is fitted to the numerical solution with high accuracy and its validity is confirmed by comparing the Hubble parameter obtained from the parameterization with the numerical solution. Using this framework, the behavior of cosmological parameters including the Hubble parameter, the comoving Hubble horizon, the energy density, the pressure, and the equation of state (EoS) parameter during the evolution of the universe is investigated, and a smooth transition from the contraction phase to the expansion phase without the occurrence of a singularity is shown. Next, the stability of the model is studied using an autonomous dynamical system and critical point analysis in phase space. The results show that the presented model, in addition to describing a non-singular bouncing scenario, has stable dynamic behavior and the introduced analytical parameterization can be used as a suitable tool for future analytical studies.

\end{abstract}



\maketitle

\newpage
\tableofcontents

%################################################################################
%################################################################################
\section{Introduction}
The importance of modified gravity theories in the study of the very universe is illustrated by the literature review of Odintsov et al. \cite{Odintsov-2023}. The review covers many aspects of modified gravity in cosmology and is focused on recent advances on inflationary cosmology in the modified gravity framework. It turns out that inflation as supported by theory and observation is the fundamental aspects of the evolutionary characteristic of the universe, and that modified theories of gravity play a central role in this aspect because they describe able to describe an inflation field as multi-fold due to its multiple shortcomings. Among the modified gravity theories, non-minimal gravitational theory enables one to obtain the asymmetric bounce, and perform the dynamical stability analysis Banerjee and et al, and Odintsov and et al. \cite{Banerjee-2022, Odintsov-2021, Odintsov-2020}. More relevantly to our work in this paper, Kerachian et al. \cite{Kerachian-2019} analyzed the dynamics of a non-minimally coupled scalar field with an unspecified positive potential in a spatially curved FRW spacetime, and studied; its dynamics  in a spatially curved FRW spacetime, defined a set of dimensionless variables which are valid for all positive, negative, and zero curvatures. The crucial point being able to define the dimensionless variables for each physical quantity so that a better interpretation of the full dynamics of the system can be attained. Furthermore the critical points were derived for a generic potential and singularities of the system, and then demonstrated that the determined critical points actually correspond to the de Sitter universe, the radiation universe, and the Milne universe.

Theories and observations suggest that during inflationary phase, the universe accelerated exponentially, and the result of these theories are observed as evidence for the acceleration of the universe in the late time. So that some evidence of the late universe can be related to type Ia supernovae and high redshift supernovae \citep{Riess-1998, Perlmutter-1999, Bennett-2003}, large scale structure \cite{Abazajian-2004}, cosmic microwave background anisotropies \cite{Spergel-2007}, Planck data \cite{Ade-2014}, and baryon acoustic oscillations \cite{Percival-2010}. During the period of inflation, the universe expanded rapidly and quickly reached a large size again. With the help of the inflation scenario, the primary problems of the universe such as the flatness and the horizon were answered. In addition to these problems, the existence of singularity is also raised as another problem of early cosmology. To analyze this problem, the universe is described as oscillatory, that is, the current universe was created due to the collapse of the previous universe. For this purpose, the Hubble parameter and the scale factor were two geometrical factors that show us the rate of expansion along the spatial direction. Now we discuss two possibilities that exist for the early time:

1) The big bang singularity occurs when the scale factor tends to zero.

2) Without reaching the singularity, the universe is increased again which represents the behavior of bouncing.

Since the scale factor can never be zero, then the space-time singularity does not occur in the early universe, as a result, the bounce scenario becomes stronger, i.e., when the scale factor reaches its minimum size, the universe begins to grow. This expression mathematically means that the slope of the scale factor or the derivative of the scale factor is zero at the mentioned point, so we call this point the bounce point. In fact, before reaching the bounce point, the previous universe contracts, and after passing this point, our new universe begins to grow again. While the size of the scale factor is decreasing at the end of the previous universe period, we can say $\dot{a} < 0$, and while the size of the scale factor is increasing at the beginning of the current universe, we can say $\dot{a} > 0$. From what was said, it can be concluded that the bounce occurs when the Hubble parameter is zero, and for the collapse of the previous universe and the beginning of the current universe is $H < 0$ and $H > 0$, respectively \cite{Cai-2014, Brandenberger-2012, Cai-2011, Sadeghi-2009, Singh-2018, Caruana-2020, Minas-2019, Skugoreva-2020, Amani-2016, Nojiri-2016, Cai-2013, Battefeld-2015, Sadeghi-2010, Lohakare-2022, Ilyas-2021, Ijjas-2018, Myrzakulov-2014, Shabani-2018, Brandenberg-2017, Bamba-2014, Tripathy-2019, Nojiri-2017, Brizuela-2010, Martin-Benito-2021, Sato-2018, Piao-2005, Cai-2009, Gordon-2003, Chattopadhyay-2017, Kerlick-1976, Alberti-2020, Novello-2008, Makarenko-2017, Brandenberger-2012a, Cai-2013a, Agrawal-2023, Sadatian-2024, Battista-2021, Assolohou-2012}.

In what follows, we discuss the role of the scalar field in the early universe, which plays a key role in the description of inflationary cosmology, and it is stated that inflation is controlled by a scalar field called the inflaton field. In standard cosmology, the corresponding action is written in terms of two terms of kinetic energy and potential along with their coupling with the curvature term for the early universe. Therefore, the influence of the scalar field in the late universe and the early universe is related to the quintessence field and the inflaton field, respectively. In this case, if a non-minimal coupling between the curvature and the quintessence field is created, we will have the description of accelerated expansion \cite{Kamenshchik-2001, Tsujikawa-2013}, and if a non-minimal coupling between the curvature and the inflaton field is created, we will have the description of inflationary cosmology \cite{Lidsey-1991, Amendola-1993, Shokri-2021}. Also, some authors investigated another model with a newer approach called non-minimal derivative coupling \cite{Capozziello-2000, Sushkov-2009, Mostaghel-2017}. Some other authors developed the non-minimal derivative coupling model as a non-minimal kinetic coupling model, so that in this model there is a coupling between the scalar field and the kinetic energy, which means that the universe evolution scenario is described by the dominance of coupling term. The non-minimal kinetic coupling theory is a form of scalar tensor theory with a scalar field with minimal coupling to gravity. Some authors explored the theory for various aspects of cosmology in the early universe and even in the late universe \cite{Dress-1985, Granda-2010a, Granda-2010b, Granda-2010c, Granda-2011, Sushkov-2012, Darabi-2015, Atazadeh-2015, Goodarzi-2022, Clifton-2012}.

Although the non-minimal kinetic coupling theory has been able to describe various cosmic behaviors in explaining the early and late universe, studying bounce scenarios in this framework, especially from the perspective of background dynamics evolution and model stability, still requires further investigation. The main motivation of this research is to study the evolution of a non-singular bouncing universe within the framework of this theory and to investigate whether the field equations can directly determine the evolution of the universe without any prior assumption for the scale factor. Therefore, first, the Friedmann equations and the field equation are solved simultaneously and numerically so that the time evolution of the scale factor, the scalar field, and other background quantities can be obtained directly from the model dynamics. Then, in order to provide an explicit form that can be used in subsequent studies, an analytical parameterization for the scale factor is introduced that reproduces the numerical solution with high accuracy. Finally, to investigate the qualitative behavior and stability of the model, the corresponding dynamical system is introduced and the structure of critical points and phase space paths are studied.

The structure of this paper is organized as follows:\\
In Section \ref{II}, the theoretical foundations of gravity with non-minimal kinetic coupling, the model action and the corresponding field equations are introduced. In Section \ref{III}, the bounce conditions and the evolution behavior of the universe within the framework of this model are investigated. In Section \ref{IV}, the coupled cosmological equation system is solved numerically and the evolution of the scale factor, scalar field, Hubble parameter, comoving Hubble horizon, energy density, pressure and equation of state parameter are analyzed. In Section \ref{V}, an analytical parameterization is presented to represent the numerical solution of the universe background and its validity is evaluated by comparing it with the Hubble parameter obtained from the numerical solution. In Section \ref{VI}, the dynamic behavior and stability of the model are investigated using an autonomous dynamical system, critical points and phase space paths. Finally, the main results of the research are summarized in Section \ref{VII}.

%################################################################################
\section{Foundations of Non-Minimal Kinetic Coupling Gravity}\label{II}
We start the Friedmann-Robertson-Walker (FRW) metric in the following form,
\begin{equation}\label{ds2}
 ds^2 = -dt^2 + a(t)^2 \left(\frac{dr^2}{1- kr^2} + r^2 ( d\theta^2 + \sin^2\theta ~d\phi^2) \right),
\end{equation}
where $a(t)$ is scale factor, and $k=0,+1,-1$ implies the flat, close and open universe,
respectively. The matter energy-momentum tensor of a perfect fluid is
$T^\mu_\nu=$ diag($-\rho,p,p,p)$. In that case, the standard equations of Friedmann are as
\begin{subequations}\label{fried1}
\begin{eqnarray}
& 3 H^2 = \kappa^2 \rho, \\
&2 \dot{H} + 3 H^2 = -\kappa^2 p,
\end{eqnarray}
\end{subequations}
where $\kappa^2 = 8 \pi G$, and $H = \dot{a} / a$ is the Hubble parameter. We are going to start with the following  action \cite{Granda-2010a, Granda-2010b, Granda-2010c, Granda-2011}, so that one has an interaction between the inflaton scalar field, $\phi$, with Ricci scalar, $R$, and Ricci tensor, $R_{\mu \nu}$,
\begin{equation}\label{action}
S = \int d^{4}x\sqrt{-g}\Big[\frac{1}{2 \kappa^2} R-\frac{1}{2}\partial_{\mu}\phi\partial^{\mu}\phi-
\frac{1}{2} \xi R \left(F(\phi)\partial_{\mu}\phi\partial^{\mu}\phi\right) - \frac{1}{2} \eta R_{\mu\nu}\left(F(\phi)\partial^{\mu}\phi\partial^{\nu}\phi\right) - V(\phi)\Big],
\end{equation}
where $\xi$ and $\eta$ are the coupling constants of dimensionless, $F(\phi)$ and $V(\phi)$ are an arbitrary functions of the scalar field and the system potential, respectively. We note that the inflaton field is an auxiliary scalar field that causes cosmic inflation in the early universe.

Thus, by taking variation of action (\ref{action}) with respect to the metric, yields
\begin{equation}\label{rmu}
G_{\mu \nu} \equiv R_{\mu\nu}-\frac{1}{2}g_{\mu\nu}R=\kappa^2\left[T_{\mu\nu}^{\phi}+T_{\mu\nu}^{\xi}+T_{\mu\nu}^{\eta}\right],
\end{equation}
where $T_{\mu\nu}^{\phi}$, $T_{\mu\nu}^{\xi}$ and $T_{\mu\nu}^{\eta}$ are the usual
energy-momentum tensor for scalar field, $\phi$, the minimally coupled terms, $\xi$ and $\eta$, respectively.
The aforesaid energy-momentum tensors are written in the following form
\begin{subequations}\label{tphi}
\begin{eqnarray}
&T_{\mu\nu}^{\phi} = \nabla_{\mu}\phi\nabla_{\nu}\phi-\frac{1}{2}g_{\mu\nu}\nabla_{\lambda}\phi\nabla^{\lambda}\phi
-g_{\mu\nu}V(\phi),\label{tphi-1}\\
& T_{\mu\nu}^{\xi} = \xi\Big[\left(R_{\mu\nu}-\frac{1}{2}g_{\mu\nu}R\right)\left(F(\phi)\nabla_{\lambda}\phi\nabla^{\lambda}\phi\right)+
g_{\mu\nu}\nabla_{\lambda}\nabla^{\lambda}\left(F(\phi)\nabla_{\gamma}\phi\nabla^{\gamma}\phi\right) \notag\\
& -\frac{1}{2}(\nabla_{\mu}\nabla_{\nu}+\nabla_{\nu}\nabla_{\mu})\left(F(\phi)\nabla_{\lambda}
\phi\nabla^{\lambda}\phi\right)+R\left(F(\phi)\nabla_{\mu}\phi\nabla_{\nu}\phi\right)\Big],\label{tphi-2}\\
&T_{\mu\nu}^{\eta} = \eta\Big[F(\phi)\left(R_{\mu\lambda}\nabla^{\lambda}\phi\nabla_{\nu}\phi+
R_{\nu\lambda}\nabla^{\lambda}\phi\nabla_{\mu}\phi\right)-\frac{1}{2}
g_{\mu\nu}R_{\lambda\gamma}\left(F(\phi)\nabla^{\lambda}\phi\nabla^{\gamma}\phi\right) \notag\\
&-\frac{1}{2}\left(\nabla_{\lambda}\nabla_{\mu}\left(F(\phi)\nabla^{\lambda}\phi\nabla_{\nu}\phi\right)+
\nabla_{\lambda}\nabla_{\nu}\left(F(\phi)\nabla^{\lambda}\phi\nabla_{\mu}\phi\right)\right) \notag\\
& +\frac{1}{2}\nabla_{\lambda}\nabla^{\lambda}\left(F(\phi)\nabla_{\mu}\phi\nabla_{\nu}\phi\right)+
\frac{1}{2}g_{\mu\nu}\nabla_{\lambda}\nabla_{\gamma}\left(F(\phi)\nabla^{\lambda}\phi\nabla^{\gamma}\phi\right)\Big].\label{tphi-3}
\end{eqnarray}
\end{subequations}

In order to obtain the equation of motion, we take variation of action with respect to the scalar field and earn as
\begin{eqnarray}\label{eqmotion}
&-\frac{1}{\sqrt{-g}}\partial_{\mu} \Big[\sqrt{-g}\left(\xi R F(\phi)\partial^{\mu}\phi+\eta R^{\mu\nu}F(\phi)\partial_{\nu}\phi+
\partial^{\mu}\phi\right)\Big] + \frac{dV}{d\phi}\notag\\
&+\frac{dF}{d\phi}\left(\xi R\partial_{\mu}\phi\partial^{\mu}\phi+\eta R_{\mu\nu}\partial^{\mu}\phi\partial^{\nu}\phi\right) = 0.
\end{eqnarray}

Now by inserting Eqs. \eqref{tphi} into Eq. \eqref{rmu}, we obtain the modified form of the Friedmann equations, and applying the restriction on $\xi$ and $\eta$ given by $\eta + 2\xi = 0$, the third and forth terms of the action \eqref{action} represent a coupling the scalar field with the Einstein tensor, $G_{\mu \nu}$, that becomes $\xi G_{\mu \nu} \left(F(\phi)\partial^{\mu}\phi\partial^{\nu}\phi\right)$. In that case, by using FRW metric \eqref{ds2} we earn the modified Friedmann equations and the field equation in the following form
\begin{subequations}\label{fried2}
\begin{eqnarray}
&3 H^2 = \kappa^2 \left(\frac{1}{2}\dot{\phi}^2+V(\phi)+9\xi H^2F(\phi)\dot{\phi}^2\right), \label{fried2-1}\\
&3 H^2+2 \dot{H} = -\kappa^2 \left(\frac{1}{2}\dot{\phi}^2-V(\phi)-\xi\left(3H^2+2\dot{H}\right)F(\phi)\dot{\phi}^2-2\xi H \left(2F(\phi)\dot{\phi}\ddot{\phi}+\frac{dF}{d\phi}\dot{\phi}^3\right)\right),\label{fried2-2}\\
&\ddot{\phi}+3H\dot{\phi}+\frac{dV}{d\phi}+3\xi H^2\left(2F(\phi)\ddot{\phi}+\frac{dF}{d\phi}\dot{\phi}^2\right)
+18\xi H^3F(\phi)\dot{\phi}+12\xi H\dot{H}F(\phi)\dot{\phi}=0.\label{fried2-3}
\end{eqnarray}
\end{subequations}

From Eqs. \eqref{fried1}, \eqref{fried2-1}, and \eqref{fried2-2} we write down the energy density and pressure as
\begin{subequations}\label{fried3}
\begin{eqnarray}
&\rho = \frac{1}{2}\dot{\phi}^2+V(\phi)+9\xi H^2F(\phi)\dot{\phi}^2, \label{fried3-1}\\
&p = \frac{1}{2}\dot{\phi}^2-V(\phi)-\xi\left(3H^2+2\dot{H}\right)F(\phi)\dot{\phi}^2-2\xi H \left(2F(\phi)\dot{\phi}\ddot{\phi}+\frac{dF}{d\phi}\dot{\phi}^3\right),\label{fried3-2}
\end{eqnarray}
\end{subequations}
where $\rho$ and $p$ are depend on the Hubble parameter, the scalar field, their time derivatives, and the system potential. It should be note that in the absence of coupling constant, $\xi$, we reach to standard inflaton model that when the potential energy of the inflation is larger than its kinetic energy, the negative pressure appears, so this issue can confirm the existence of potential energy in early cosmology. But in this research, in addition to potential energy being greater than kinetic energy, in the presence of coupling constant, $\xi$, the third and the forth terms of Eq. \eqref{fried3-2} with the condition that parameters $H$, $\dot{H}$, $\phi$, $\dot{\phi}$, $\ddot{\phi}$, $F(\phi)$ and $dF/d\phi$ are positive lead to a more negative pressure, which is a convincing description for inflationary cosmology. In that case, the current model is expected to satisfy the positivity of the mentioned parameters, which will be considered in the next sections. 

Therefore, the field EoS parameter reads
\begin{equation}\label{eosphi}
\omega=\frac{p}{\rho}=\frac{\frac{1}{2}\dot{\phi}^2-V(\phi)-\xi\left(3H^2+2\dot{H}\right)F(\phi)\dot{\phi}^2 - 2\xi H\left(2F(\phi)\dot{\phi}\ddot{\phi}+\frac{dF}{d\phi}\dot{\phi}^3\right)}{\frac{1}{2}\dot{\phi}^2+V(\phi)+9\xi H^2F(\phi)\dot{\phi}^2},
\end{equation}
where the EoS is depend on the scalar field and its derivatives with respect to cosmic time, the potential, and the Hubble parameter. In next section, we will study the behavior of bouncing universe.

%$$$$$$$$$$$$$$$$$$$$$$$$$$$$$$$$$$$$$$$$$$$$$$$$$$$$$$$$$$$$$$$$$$$$$$$$$$$$$$$$$$$$$$$$$$
%$$$$$$$$$$$$$$$$$$$$$$$$$$$$$$$$$$$$$$$$$$$$$$$$$$$$$$$$$$$$$$$$$$$$$$$$$$$$$$$$$$$$$$$$$$
\section{Bounce Conditions and Cosmological Evolution}\label{III}
In this section, we describe bouncing solution by non-minimal kinetic coupled reconstruction. A successful bounce requires the necessary conditions during the different phases, i.e., the contraction phase shows that the scale factor $a(t)$ is decreasing ($\dot{a} < 0$), and in the expansion phase $\dot{a} > 0$, but at the bouncing point, $\dot{a} = 0$, and around this point $\ddot{a} > 0$. Similarly, in the bouncing cosmology, the Hubble parameter, $H$, crosses zero from contraction phase ($H < 0$) to expansion phase (${H}> 0$), and at the bouncing point $H_b = 0$ and $\dot{H}_b > 0$ in which $H_b$ is the Hubble parameter at the bouncing point. So, if a question arises here, does the energy density also become zero when the Hubble parameter in the first classical Friedmann equation becomes zero? The answer is that usually at the bounce point the value of the scale factor is minimum, in which case $\dot{a}_b = 0$ and as a result $H_b = 0$, which in the dynamical state of the universe represents the transition from contraction to expansion, not the disappearance of all energy or matter, but at this point the energy reaches a significant value and even a maximum.

Therefore, a successful bounce in the standard cosmology \eqref{fried1} has the below necessary condition around bouncing point
\begin{equation}\label{doth}
\dot{H}_b=-\frac{\kappa^2}{2}(1+\omega) \rho > 0,
\end{equation}
where $\rho + p < 0$, and this is the same statement regarding the violation of the null energy condition. So this violation helps the universe to continue contracting before the Big Bang and not go back, so that the cosmic bounce occurs and the universe continues to expand after the Big Bang.

In order to obtain the bouncing condition for non-minimal kinetic coupling, we sum Eqs. \eqref{fried3} and we obtain as
\begin{equation}\label{rhop}
\rho+p = \dot{\phi}^2+6\xi H^2 F \dot{\phi}^2 - 2 \xi \dot{H} F \dot{\phi}^2 - 2 \xi H \left(2 F \dot{\phi} \ddot{\phi} + \frac{dF}{d\phi}\dot{\phi}^3\right),
\end{equation}
where by applying the bouncing condition \eqref{doth}, and violation of the null energy condition ($\rho + p < 0$) we will have
\begin{equation}\label{con1}
\xi \dot{H}_b F_b >\frac{1}{2},
\end{equation}
where index $"b"$ is related to the bounce point. According to the bouncing theory, the big bang is the result of the beginning of an expansion period after a contraction period, so that, $t_b$ is the point time between contraction and expansion periods, the so-called bouncing point time.

%#################################################################################
%#################################################################################

\section{Numerical Solution of the Coupled Cosmological Equations}\label{IV}

Investigating the dynamics of a physical system such as inflationary cosmology includes terms of kinetic energy and potential energy, which the term of kinetic energy has its own common form, but the term of potential energy is very important because it determines the properties of the inflationary phase. Since the shape of the potential is very sensitive and dependent on a specific model, the chosen shape for the inflationary potential should be such that it can describe some features of inflation. Therefore, we consider the inflaton field potential as follows
\begin{equation}\label{pot1}
V(\phi)=\frac{\textrm{v}_0}{\cosh({\lambda \phi})},
\end{equation}
where $\textrm{v}_0$ and $\lambda$ are constant. This potential has been studied in many cosmological models, including scalar fields, tachyon fields, teleparallel gravity, and brane--world gravity \cite{Sahni-2000, Singh-2003, Geng-2012, Steer-2004, Leblond-2003}. This type of potential, which is associated with scalar fields, determines how the field evolves over time and how it affects the expansion of the early universe. In this case, the corresponding potential shape implies a scalar field that can roll down from  the hilltop of near the origin and end up in the inflation period after passing through a steep region. Therefore, it could have implications for the production of perturbations in the early universe and the subsequent formation of large-scale structures. In what follows, considering the effect of the third and fourth terms of action \eqref{action} on the corresponding potential, it is necessary to consider the form of function $F(\phi)$ as a power law to make a balance between potential and non-minimal kinetic gravity. However, we consider $F(\phi) = \sum\limits_{n=0} c_n \phi^{2 n}$ in which $c_0=1$, $c_1=c$, and $c_{n \geq 2}=0$.

%#################################################################################
\subsection{Evolution of the Scale Factor and Scalar Field}

After determining the form of the potential function \(V(\phi)\) and the non-minimal coupling function \(F(\phi)\), the dynamics of the model is completely determined by the Friedmann equations and the scalar field equation, that is, the Eqs. \ref{fried2-1} and \ref{fried2-3}. This system consists of two coupled nonlinear differential equations for two unknown functions \(a(t)\) and \(\phi(t)\). Due to the non-minimal interaction between the kinetic energy of the scalar field and the Einstein tensor, the resulting equations have a highly nonlinear structure and it is not possible to find a closed analytical solution for them, at least for the present choice of \(V(\phi)\) and \(F(\phi)\). Therefore, the study of the dynamics of the model is based on the direct numerical solution of this system.

For this purpose, the equations are solved simultaneously and using appropriate initial conditions to obtain the time evolution of both functions \(a(t)\) and \(\phi(t)\). The important point is that in this process no initial assumptions are made for the scale factor or the scalar field, rather, both quantities are extracted directly from the numerical solution of the field equations. Therefore, the cosmological behavior of the model is completely determined by the structure of the theory, the choice of the potential function, the coupling function, and the initial conditions.\\
Since the dynamic behavior of the model is sensitive to the values of the free parameters, the parameter space of the model is first investigated numerically. The aim of this investigation is to find a region of the parameter space that satisfies the necessary conditions for the occurrence of a regular bouncing universe. Specifically, the parameters are chosen such that the Hubble parameter transitions from negative values in the contraction phase to positive values in the expansion phase, such that at the bounce point the conditions $H=0$ and $\dot{H} > 0$ hold, while the evolution of the universe remains numerically stable.

\begin{figure}[htbp]
\begin{center}
\subfigure[]{\includegraphics[scale=.35]{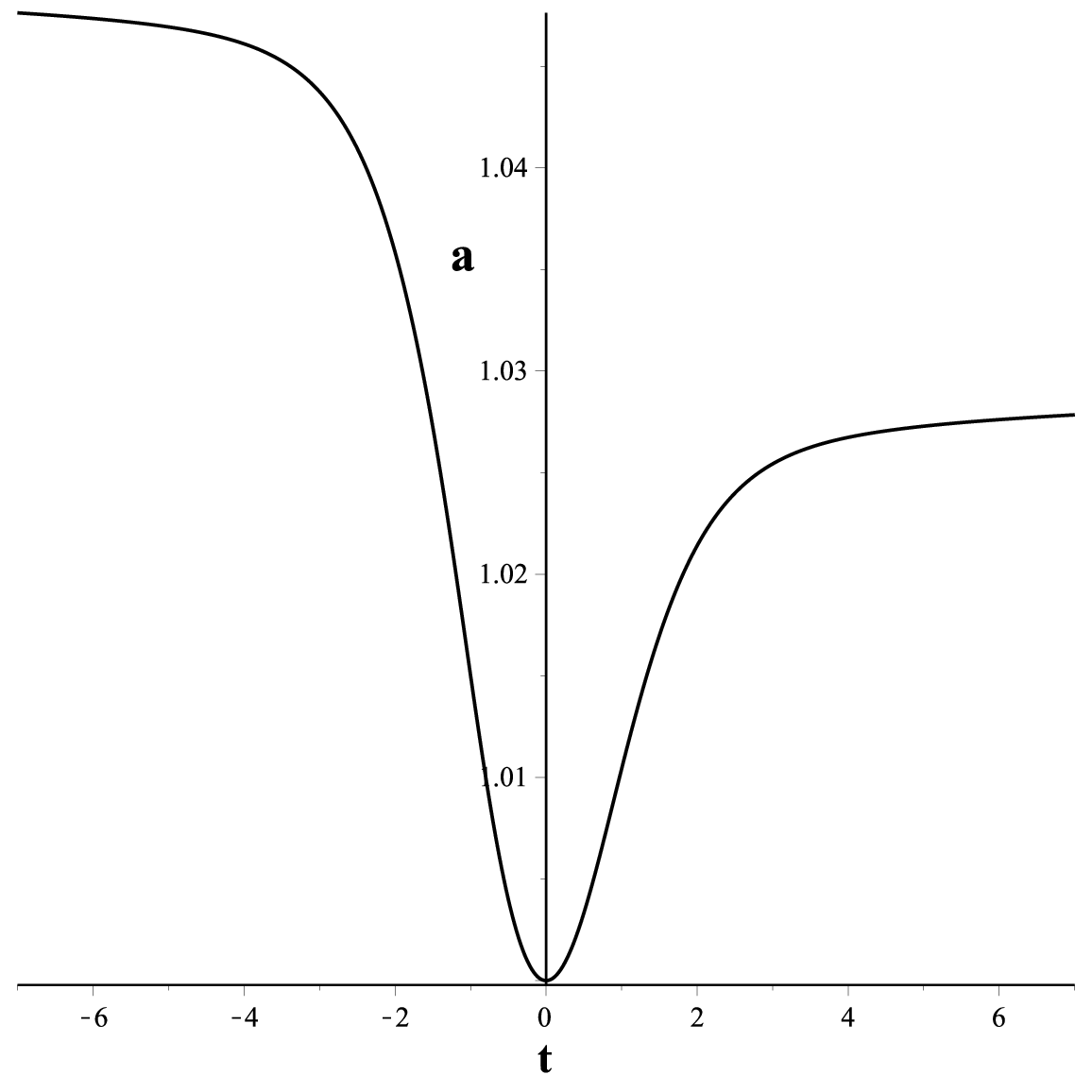} \label{fig1}}
\subfigure[]{\includegraphics[scale=.35]{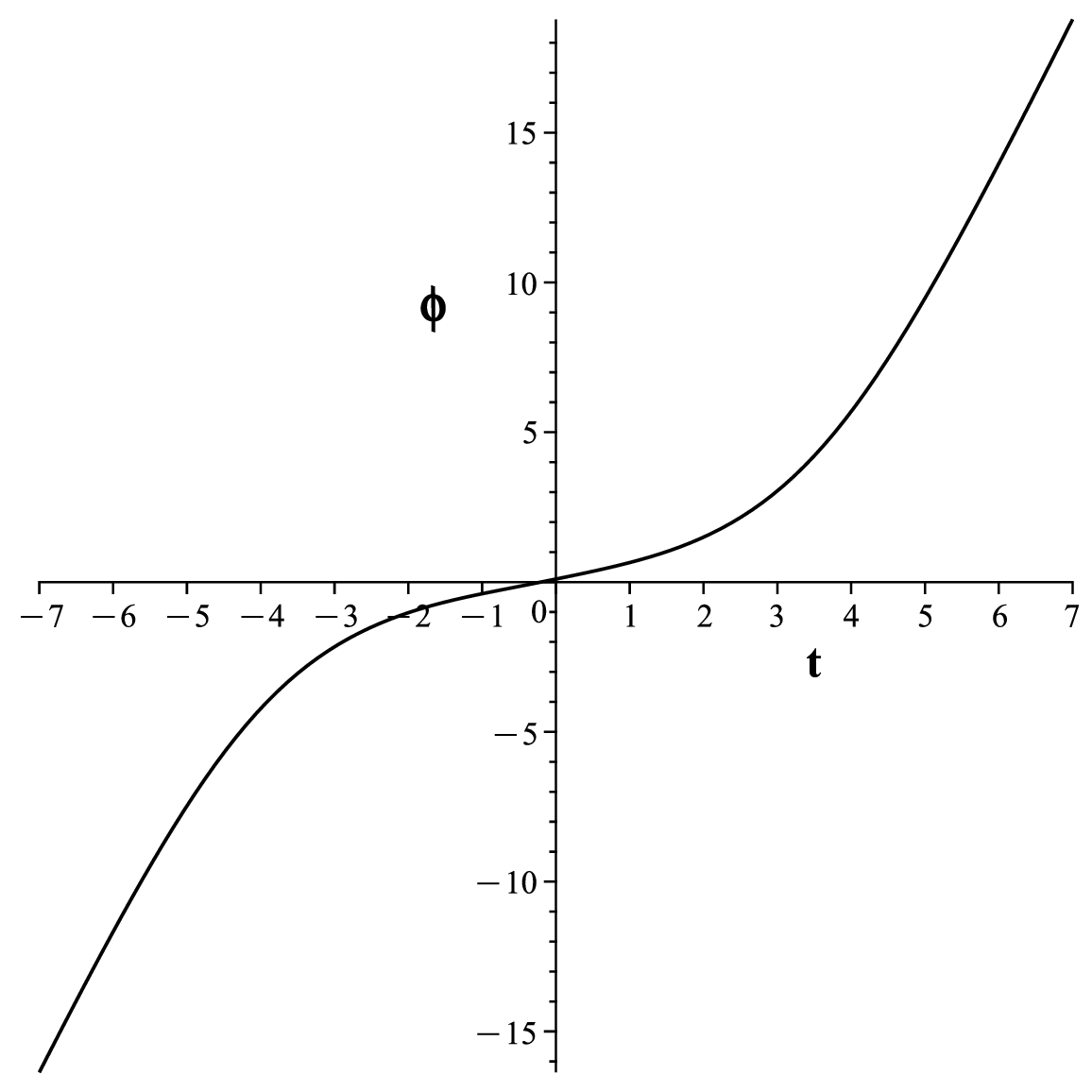} \label{field}}
\caption{Graphs of the scale factor and the scalar field in terms of cosmic time by $\xi=20$, $\textrm{v}_0=12.5$, $c=1$, and $\lambda=0.2$ and with initial conditions $a(0)=1$, $\dot{a}(0)=0$, $\phi(0)=0.1$, $\dot{\phi}(0)=0.5$.} \label{fig1field}
\end{center}
\end{figure}

After examining the parameter space, the set of values $\xi=20$, $\textrm{v}_0=12.5$, $c=1$, and $\lambda=0.2$ along with the initial conditions $a(0)=1$, $\dot{a}(0)=0$, $\phi(0)=0.1$, and $\dot{\phi}(0)=0.5$ is chosen as the set that produces a regular and physically plausible bounce solution.\\
The numerical solution of the system of equations determines the time evolution of both functions \(a(t)\) and \(\phi(t)\) simultaneously. The results of this numerical solution are presented in Figs. \ref{fig1} and \ref{field}, respectively. Fig. \ref{fig1} shows the evolution of the scale factor, while Fig. \ref{field} shows the time behavior of the scalar field corresponding to the same numerical solution. These two diagrams are in fact the direct output of the system of field equations and form the basis of all subsequent analyses in the paper.

It can be seen from Fig. \ref{fig1} that before the bounce point, the derivative of the scale factor is negative and the universe is in a contraction phase, while after passing the bounce point, the sign of the derivative changes and it enters an expansion phase. Also, one of the notable features of the numerical solution is the asymmetry of the evolution before and after the bounce, such that the contraction rate before the bounce is greater than the expansion rate after it. This behavior is a direct consequence of the interaction between the scalar field and the non-minimal coupling term, and shows that the two branches before and after the bounce are not dynamically equivalent.

In contrast, Fig. \ref{field} shows the time evolution of the scalar field obtained simultaneously with the scale factor from the numerical solution of the system of equations. In this paper, the scalar field is used solely through its numerical values in the calculation of cosmological quantities, so, unlike the scale factor, there is no need to introduce an independent analytical approximation or parameterization for \(\phi(t)\).

%#################################################################################
\subsection{Evolution of the Hubble Parameter}

\begin{figure}[htbp]
\begin{center}
\includegraphics[scale=.35]{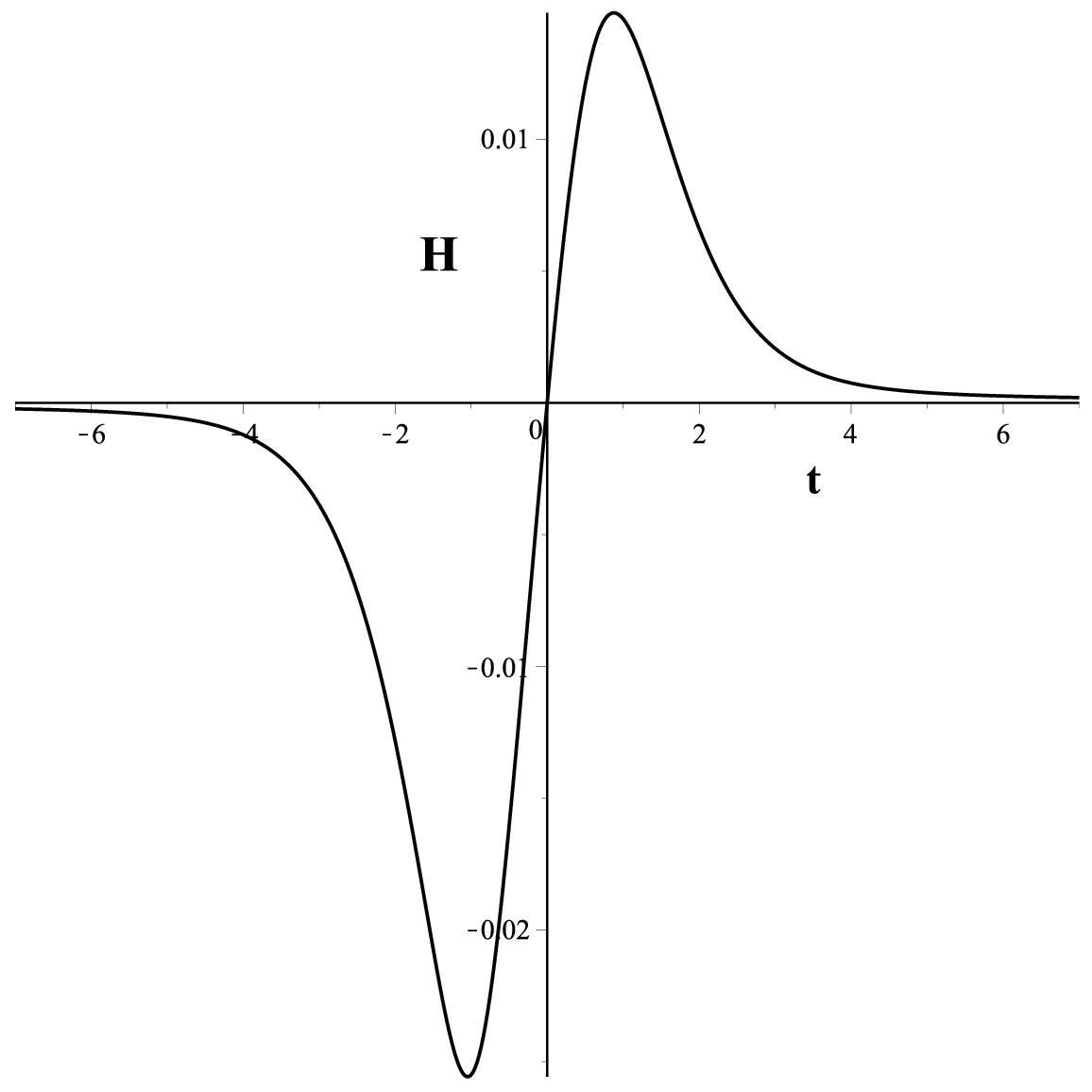}
\caption{Graph of the Hubble parameter in terms of cosmic time by $\xi=20$, $\textrm{v}_0=12.5$, $c=1$, and $\lambda=0.2$ and with initial conditions $a(0)=1$, $\dot{a}(0)=0$, $\phi(0)=0.1$, $\dot{\phi}(0)=0.5$.}\label{fig2}
\end{center}
\end{figure}

One of the most important cosmological quantities for investigating the nature of a bouncing universe is the Hubble parameter, which determines the expansion or contraction rate of the universe. After obtaining the numerical solution of the scale factor and the scalar field from the system of Friedmann equations and the field equation, the Hubble parameter is also calculated directly from the same numerical solution using the relation \(H(t) = \dot{a}/a\). Therefore, Fig. \ref{fig2} is a direct representation of the dynamics obtained from the numerical solution of the field equations and is not based on any additional assumptions or independent analytical approximations.

As can be seen in Fig. \ref{fig2}, the Hubble parameter has negative values at early times, indicating that the universe is in a contraction phase. As the bounce time approaches, the value \(H(t)\) increases continuously and reaches the value \(H = 0\) at the bounce moment. After passing this point, the Hubble parameter becomes positive and the universe enters the expansion phase. In addition, the positive slope of the curve at the point of crossing the origin shows that the condition \(\dot{H} > 0\) also holds, therefore, both basic conditions necessary for the realization of a regular bouncing universe are satisfied simultaneously.

From a physical point of view, this behavior shows that the evolution of the universe takes place without passing through an initial singularity. In this model, the universe first enters a period of contraction and its volume continuously decreases, but before reaching a singularity, due to the dynamics arising from the non-minimal interaction of the scalar field with the geometry of space-time, the contraction stops and a smooth transition (bounce) to the expansion phase occurs. As a result, the evolution path of the universe is continuously transferred from the contraction branch to the expansion branch, without the dynamical quantities diverging.

Also, Fig. \ref{fig2} shows that the behavior of the Hubble parameter on both sides of the bounce point is not completely symmetrical. This asymmetry is a reflection of the nonlinear dynamics due to the non-minimal kinetic coupling and states that the pre-bounce and post-bounce branches are not dynamically equivalent. This feature is a direct consequence of the structure of the present model, and in the following, its effect on other cosmological parameters, including energy density, pressure, EoS parameter, and dynamical systems analysis, will be investigated.

%#################################################################################
\subsection{Evolution of the Comoving Hubble Horizon}

\begin{figure}[h]
\begin{center}
\includegraphics[scale=.35]{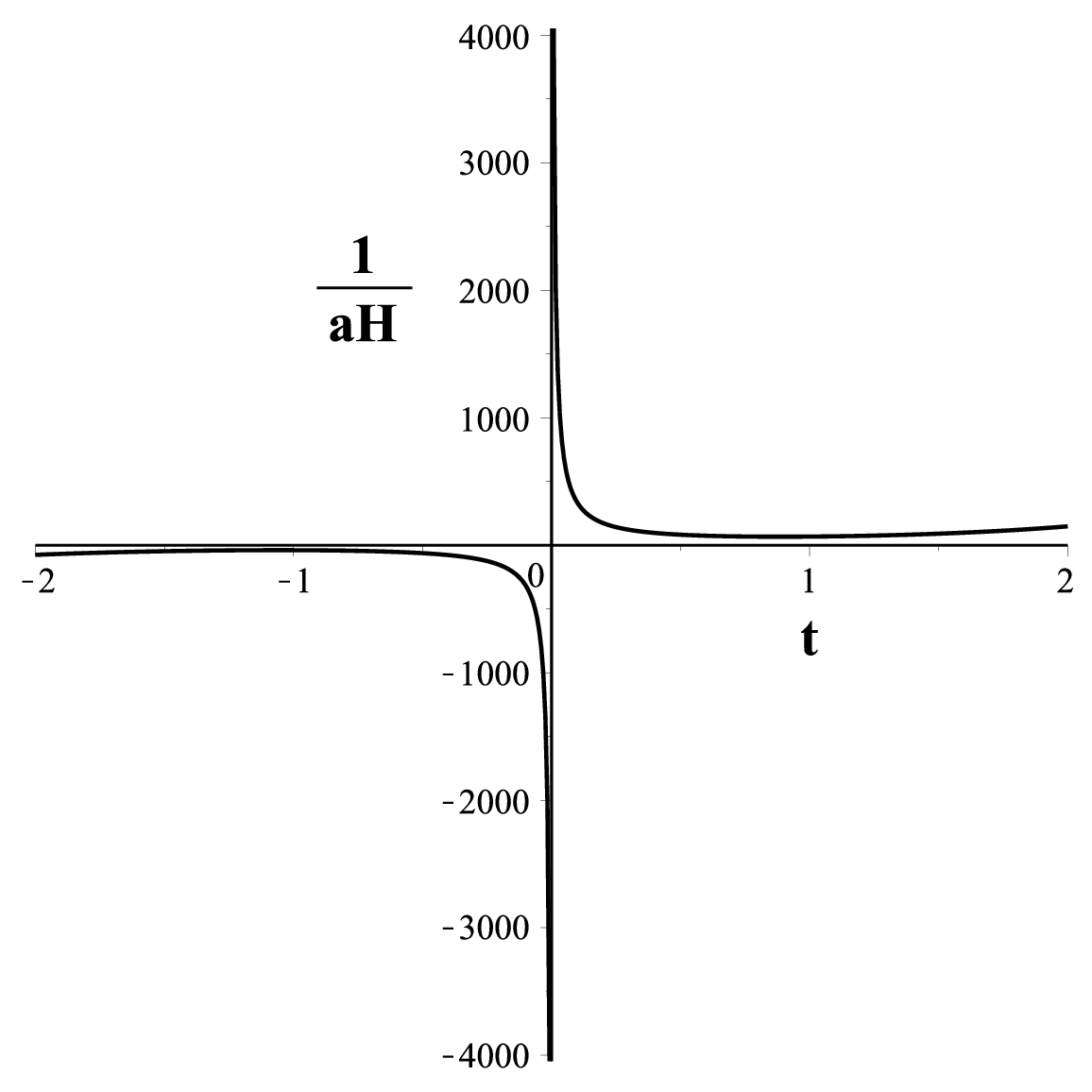}
\caption{Evolution of the comoving Hubble horizon \((aH)^{-1}\) in terms of cosmic time.}\label{inverseaH}
\end{center}
\end{figure}

After examining the behavior of the scale factor and the Hubble parameter, another important quantity that provides valuable information about the dynamical structure of the comoving universe is the comoving Hubble radius or the comoving Hubble horizon, which is defined as $(aH)^{-1}$. This quantity plays a fundamental role in studying the evolution of cosmological perturbations, because it determines whether the perturbation modes at each stage of the evolution of the universe are inside or outside the Hubble horizon. Therefore, examining the temporal behavior of \((aH)^{-1}\) is considered one of the important tests for analyzing bouncing models and comparing them with standard cosmological scenarios. Using the direct numerical solution of the Friedmann equations and the field equation, the quantity \((aH)^{-1}\) is also calculated without any additional assumptions and is shown in Fig. \ref{inverseaH} in terms of cosmic time. As can be seen, in the contraction phase \(t < 0\) the value of this quantity is negative, because the Hubble parameter has negative values. By approaching the bounce time, the value of \(H\) tends to zero and as a result \((aH)^{-1}\) mathematically diverges. This divergence is simply due to the Hubble parameter becoming zero and does not indicate the presence of any physical singularity in the model, because the scale factor and other dynamical quantities remain finite at the bounce point. After passing the bounce point, the Hubble parameter changes its sign and becomes positive, as a result, \((aH)^{-1}\) starts from very large positive values and gradually decreases as the expansion continues. This behavior indicates the continuous transition of the universe from a contraction phase to an expansion phase and is fully consistent with the conditions necessary for the realization of a non-singular bouncing universe.

From a physical point of view, the behavior of Fig. \ref{inverseaH} shows that the size of the comoving Hubble horizon changes dramatically on both sides of the bounce. Such behavior is of fundamental importance for studying the re-exit and re-entry of cosmological perturbation modes from the horizon and suggests that the present model has the expected features of a bouncing scenario in terms of horizon structure. A detailed examination of the spectrum of perturbations and the observational consequences of this behavior requires an analysis of cosmological perturbations, which is beyond the scope of the present study.

%#################################################################################
\subsection{Energy density, Pressure, and EoS Parameter}

\begin{figure}[h]
\begin{center}
\subfigure[]{\includegraphics[scale=.3]{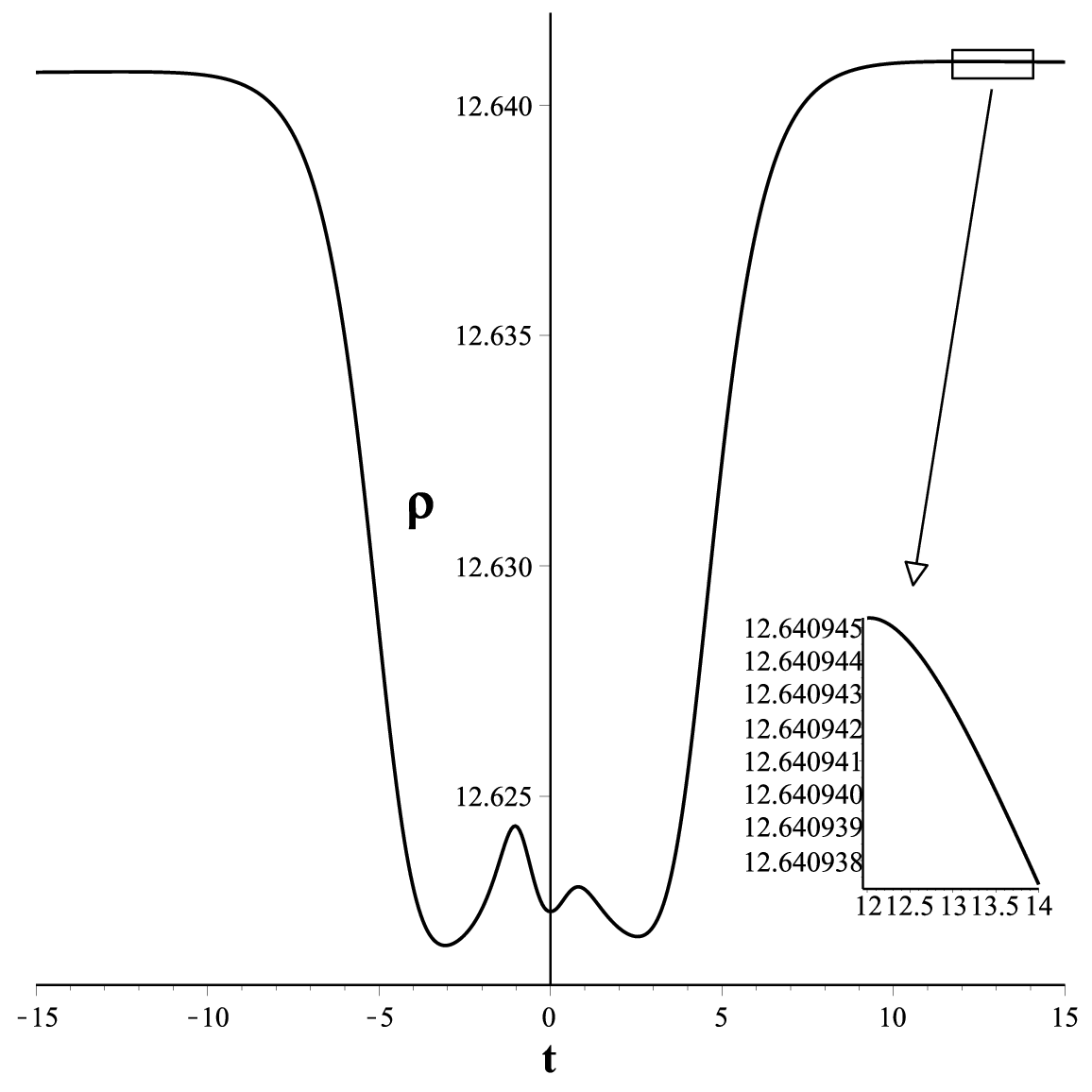}\label{rhopress-a}}
\subfigure[]{\includegraphics[scale=.3]{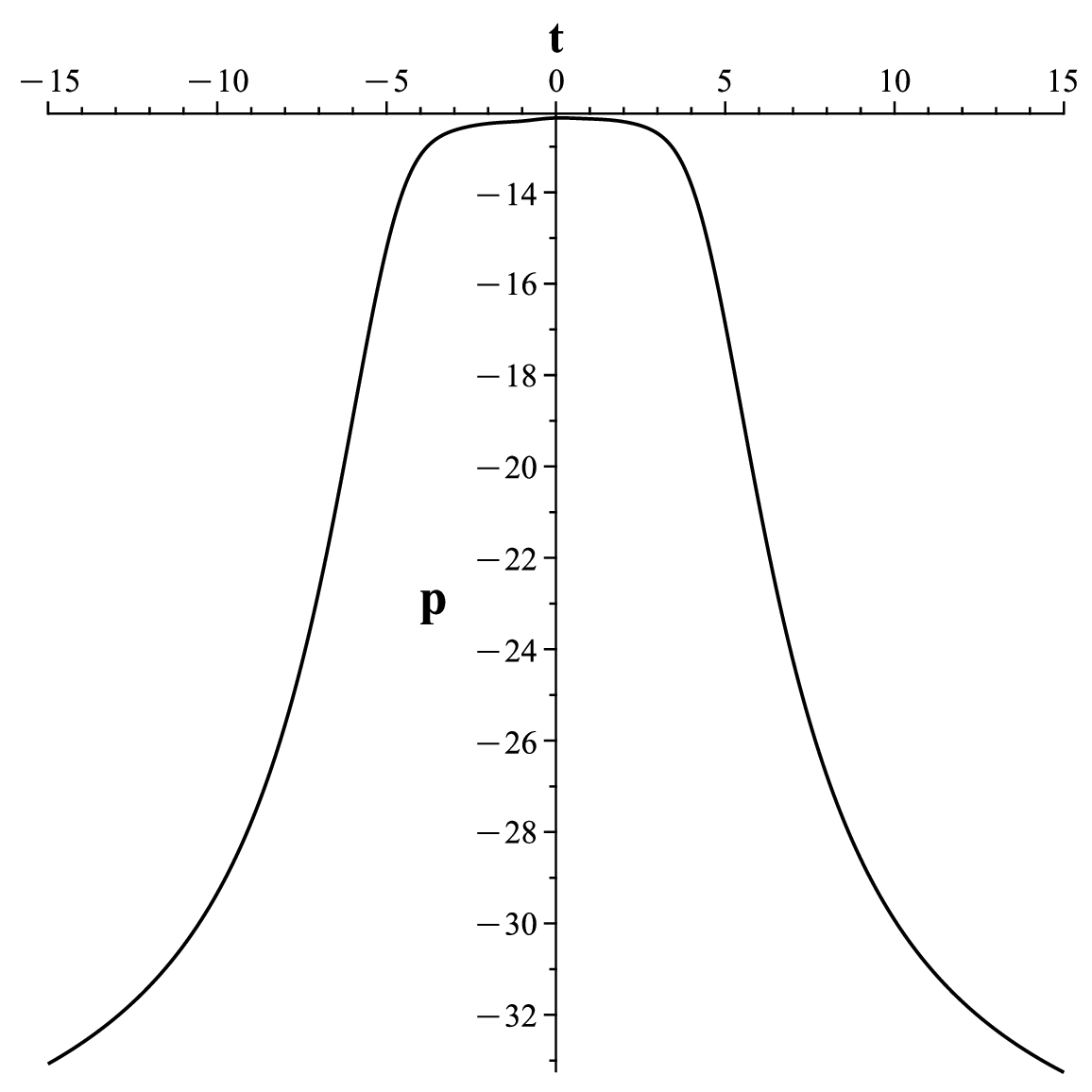}\label{rhopress-b}}
\caption{Graphs of the energy density and the pressure in terms of cosmic time by $\xi=20$, $\textrm{v}_0=12.5$, $c=1$, and $\lambda=0.2$ and with initial conditions $a(0)=1$, $\dot{a}(0)=0$, $\phi(0)=0.1$, $\dot{\phi}(0)=0.5$.}\label{rhopress}
\end{center}
\end{figure}

After determining the behavior of the scale factor, the Hubble parameter, and the comoving Hubble horizon, other cosmological quantities can now be extracted directly from the same numerical solution of the field equations. Using the Eqs. \eqref{fried3}, the energy density and the pressure have been calculated as two fundamental quantities of the dynamics of the universe, and their time variations are shown in Figs. \ref{rhopress}. Since these quantities are directly obtained from the numerical solution of the Friedmann equations and the field equation, their behavior is a direct reflection of the dynamics of the present model.

The Fig. \ref{rhopress-a} shows that the evolution of the energy density provides a global picture of the cosmological dynamics predicted by the present model and clearly distinguishes three successive regimes: the contracting phase, the bounce transition, and the expanding phase.  the energy density evolves smoothly during the contracting phase and exhibits only a slow variation while the universe remains far from the bounce. As the system approaches the bounce region, significant changes appear in it. Around the bouncing phase, the numerical solution displays a localized oscillatory behavior, indicating that the transition from contraction to expansion occurs over a finite interval rather than at an instantaneous moment. After the bounce, the energy density grows rapidly during the early expanding era and subsequently approaches a much smoother evolution with a gradual decrease toward late cosmic times. As a result, the energy density remains finite throughout the evolution of the universe and no divergence is observed around the jump; this is fully consistent with the non-singular nature of the jump scenario.

On the other hand, the pressure behavior also shows a similar dynamic evolution. According to Fig. \ref{rhopress-b}, the pressure has a relatively constant negative value in the early stages and reaches its maximum value as it approaches the jump region. After passing through the jump point, the pressure decreases again and approaches an asymptotic behavior in the later stages of the expansion. The negativity of the pressure over a significant period of the evolution of the universe indicates the role of the scalar field in creating a gravitational repulsive component that provides the necessary condition for the jump to occur and prevents the formation of a singularity.

The combination of the behavior of the energy density and pressure shows that the transition from the contraction phase to the expansion phase occurs without any discontinuity or divergence in the thermodynamic quantities. This feature is one of the important results of the non-minimal kinetic coupling in the present model and states that geometric modifications introduced into the interaction can create a smooth and non-singular cosmological evolution.

\begin{figure}[h]
\begin{center}
\includegraphics[scale=.35]{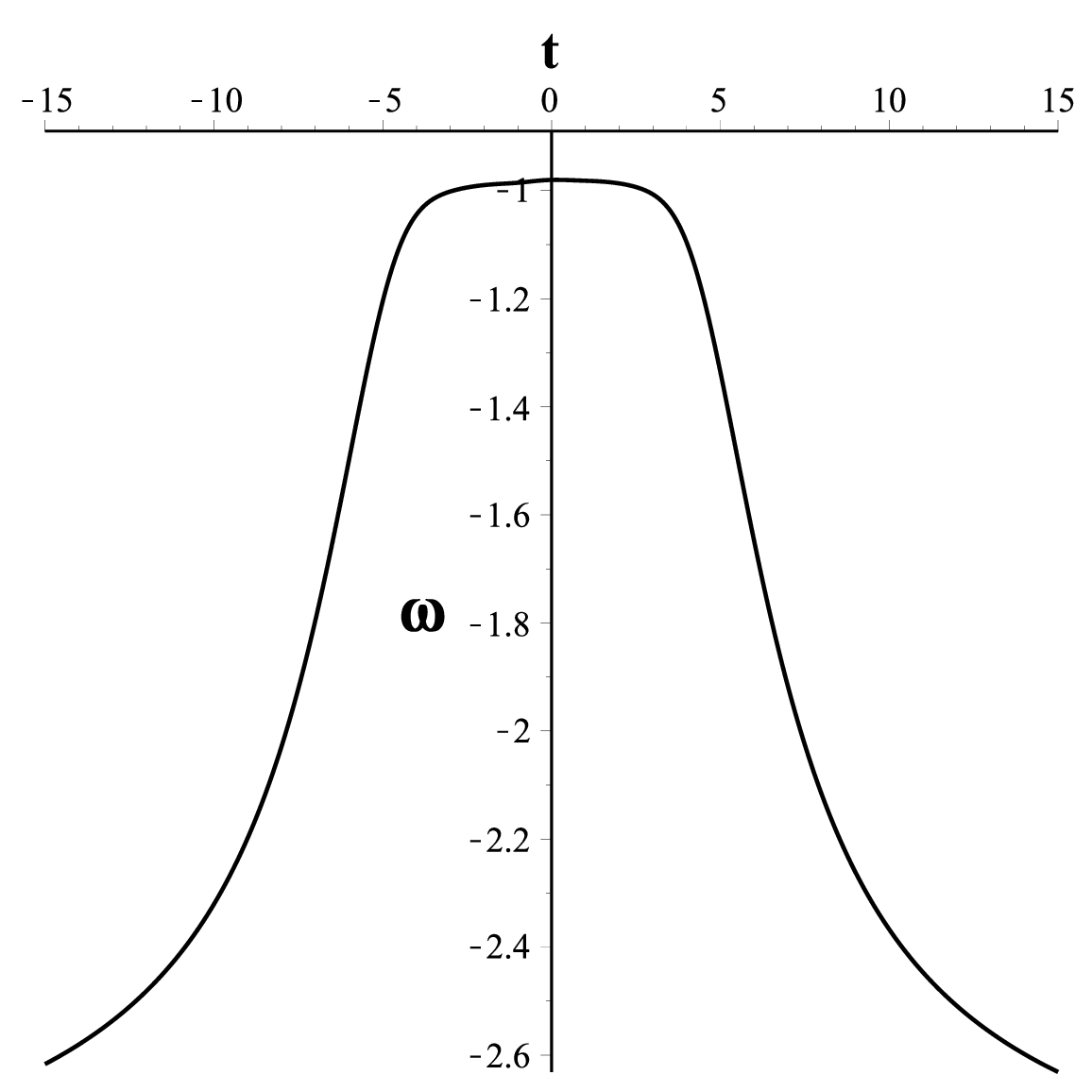}
\caption{Graph of the EoS parameter in terms of cosmic time by $\xi=20$, $\textrm{v}_0=12.5$, $c=1$, and $\lambda=0.2$ and with initial conditions $a(0)=1$, $\dot{a}(0)=0$, $\phi(0)=0.1$, $\dot{\phi}(0)=0.5$.}\label{omega}
\end{center}
\end{figure}

To examine the properties of the matter in more detail, the EoS parameter \eqref{eosphi} is also calculated based on the same numerical solution and is shown in Fig. \ref{omega}. As can be seen, the value of the EoS parameter in the early period of contraction is in the phantom region \(\omega < -1\). By approaching the transition point, its value increases and briefly enters the quintessence region \(\omega > -1\). After passing through the transition, the evolution process is reversed and, as the expansion continues, the EoS parameter returns to the phantom region. This transition between two different regimes of the EoS indicates that the scalar field, together with the non-minimal kinetic coupling, is able to create different dynamical behaviors at different epochs of the evolution of the universe. As a result, the present model not only provides the necessary conditions for the occurrence of a non-singular transition universe, but is also able to describe the effective evolution of the matter of the universe from the pre-transition period to the late epochs of the expansion in a continuous and regular.

The results of this section show that the numerical solution of the Friedmann equations and the field equation provides a coherent picture of the evolution of the cosmic background, including the behavior of the scale factor, the scalar field, the Hubble parameter, and other cosmological quantities, in the bouncing scenario with non-minimal kinetic coupling. However, since many analytical studies and subsequent applications require an explicit form of the scale factor, in the next section we present an analytical parameterization to accurately represent the numerical solution that can reproduce the background dynamics with high accuracy.

%#################################################################################
%#################################################################################
\section{Analytical Parametrization of the Numerical Background Evolution}\label{V}

In the previous section, it was shown that the numerical solution of the differential system of Friedmann and the scalar field equations determines the time evolution of the scale factor and the scalar field (as shown in Figs. \ref{fig1field}) without assuming any analytical form for \(a(t)\). Although this numerical solution is quite sufficient for investigating the cosmological properties of the model, it would be very useful to have an explicit analytical representation of the scale factor in many subsequent applications, such as studying cosmological perturbations, reconstructing potentials, thermodynamic analysis, calculating observational quantities, and comparing with other models. For this reason, in this section, we introduce an analytical parameterization to approximate the numerical solution. We emphasize that this function is not extracted from the field equations, but is merely a high precision analytical approximation to represent the numerical solution obtained from the field equations. This parameterization can be used in future  studies as a suitable alternative for numerical data. For this purpose, we consider the following function for the scale factor as
\begin{equation}\label{at2}
a(t)=\frac{A e^{2 \mu t}+B e^{\mu t}+C}{\left(D e^{\mu t}+1\right)^2},
\end{equation}
where \(A\), \(B\), \(C\), \(D\), and \(\mu\) are fitting coefficients, such that the coefficients \(A\), \(B\), \(C\), \(D\) are dimensionless and the coefficient \(\mu\) has the dimension of the inverse of time. This relation is able to reproduce the asymmetric behavior of the scale factor (as shown in Fig. \ref{fig1}) in the two phases of contraction and expansion, as well as the existence of a smooth minimum in the bounce region with very good accuracy. The important point is that the coefficients \(A\), \(B\), \(C\), \(D\), and \(\mu\) are not fundamental parameters of the theory and should not be interpreted as independent parameters of the model. These coefficients are actually effective parameters that are indirectly determined by the physical parameters of the model, \(\xi\), \(\textrm{v}_0\), \(c\), and \(\lambda\), as well as the initial conditions \(a(0)\), \(\dot{a}(0)\), \(\phi(0)\), and \(\dot{\phi}(0)\).
\begin{figure}[htbp]
\begin{center}
\includegraphics[scale=.35]{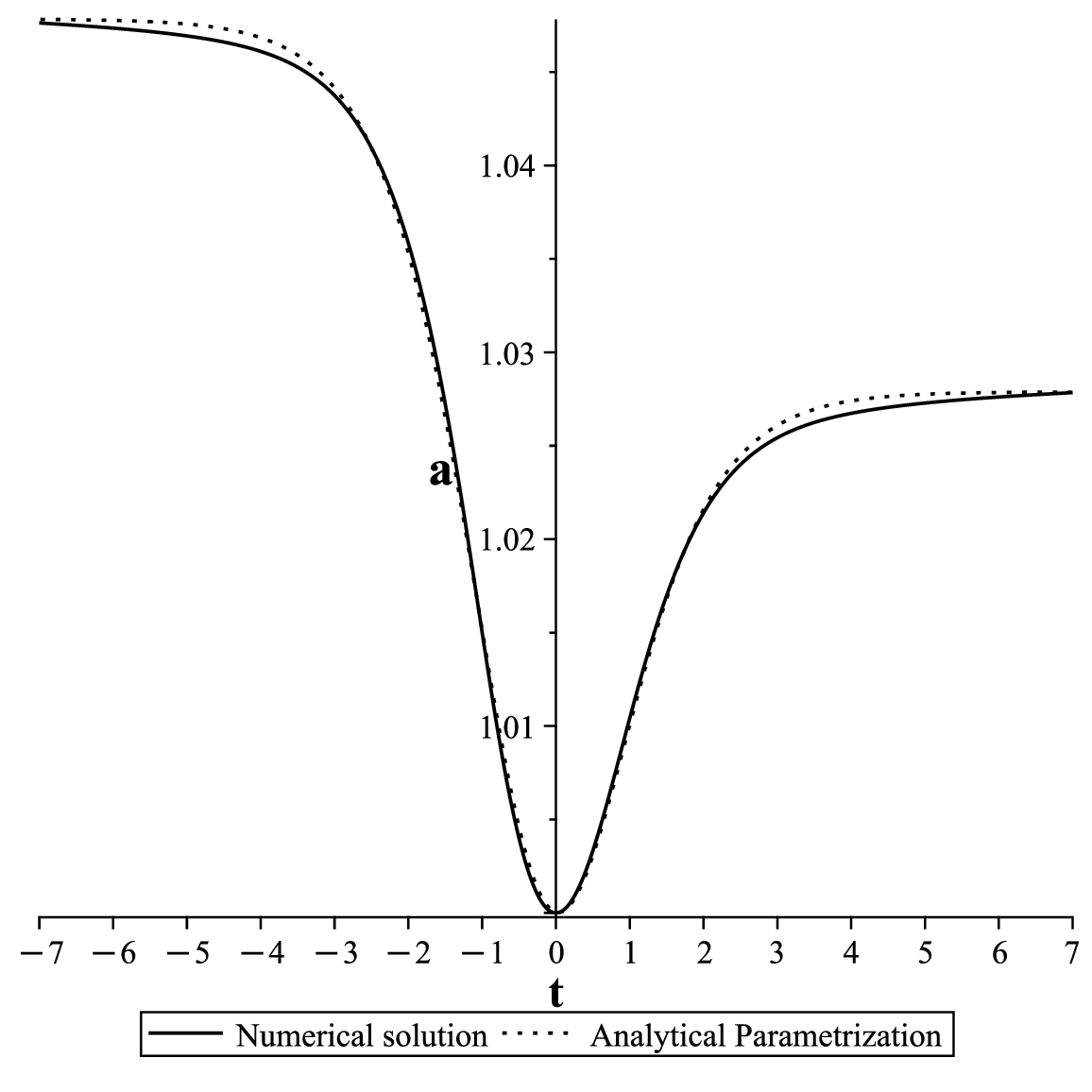}
\caption{Comparison between the numerical solution of the scale factor and its analytical parametrization.}\label{a2t-1}
\end{center}
\end{figure}

However, by fitting this function with the numerical solution obtained by solving the Friedmann equations and the field equation, the coefficient values are obtained as $A = 1.62361307169493$, $B = 2.42190962055420$, $C = 1.04783638795486$, $D = 1.25680758401826$, and $\mu = 1.33674422021756$. In Fig. \ref{a2t-1}, the numerical solution and the analytical parameterization are shown with solid and dotted lines, respectively. As can be seen, the agreement between the two curves is very good over the entire time period studied, therefore, the function \eqref{at2} can be used as a valid analytical approximation for the cosmological evolution of the present model. For this reason, in this paper the coefficients \(A\), \(B\), \(C\), \(D\) and \(\mu\) are reported only for the selected set of parameters, while extracting their explicit dependence on the fundamental parameters \(\xi\), \(\textrm{v}_0\), \(c\), and \(\lambda\) can be the subject of an independent study in the future.

\begin{figure}[h]
\begin{center}
\includegraphics[scale=.35]{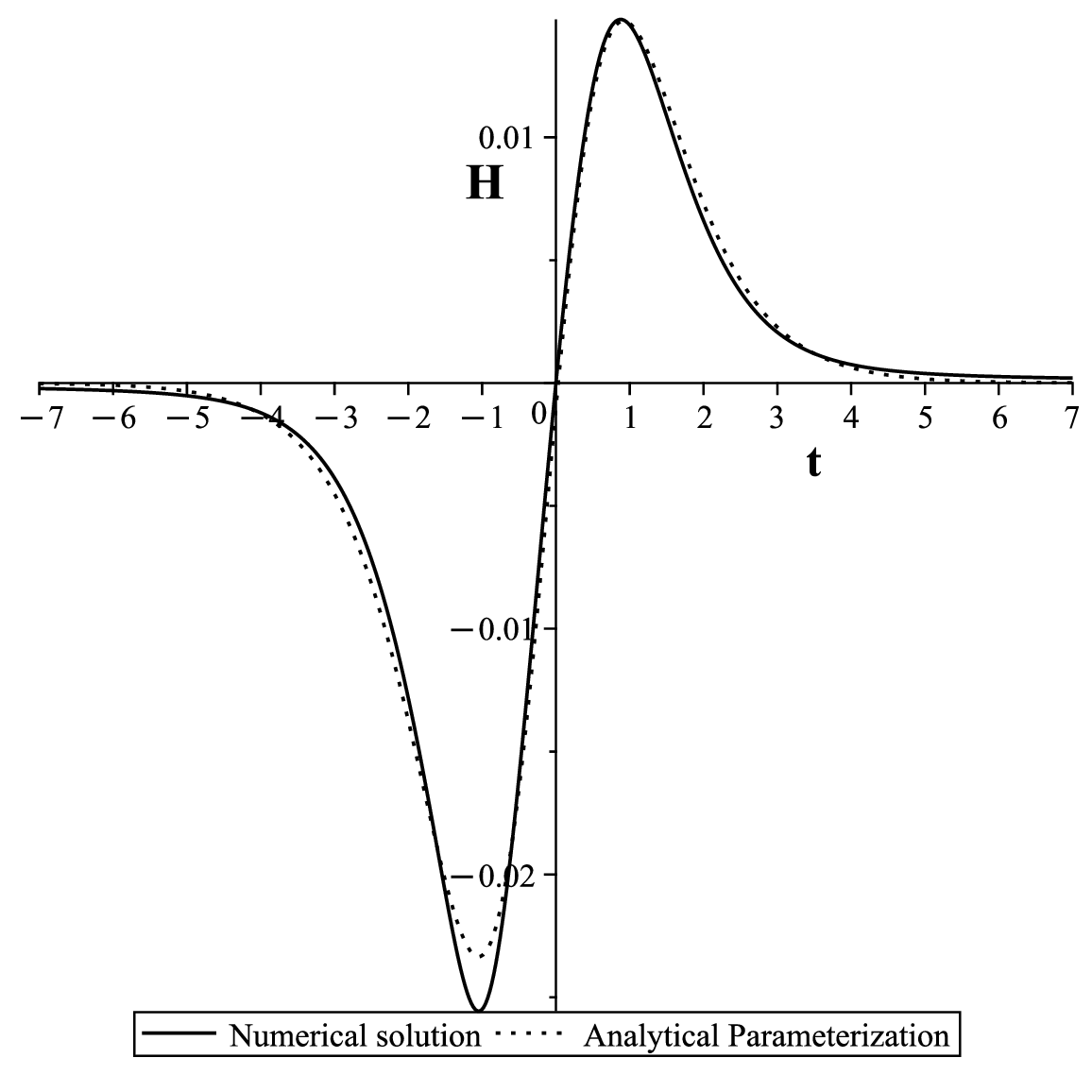}
\caption{Comparison between the numerical Hubble parameter and the Hubble parameter obtained from the analytical parametrization.}\label{H2t2}
\end{center}
\end{figure}

After introducing the analytical parameterization of the scale factor in Eq. \eqref{at2}, it is necessary to evaluate its accuracy in reproducing the model dynamics. Since the Hubble parameter is obtained directly from the time derivative of the scale factor as \(H(t)=\frac{\dot a(t)}{a(t)}\), comparing the behavior of the Hubble parameter obtained from the analytical parameterization with the Hubble parameter corresponding to the numerical solution is a suitable test for assessing the validity of this parameterization. To this end, first, using Eq. \eqref{at2}, the Hubble parameter is calculated analytically, the result of which is presented in the following form
\begin{equation}\label{Hubble1}
H = \frac{2 A \mu  \,{e}^{2 \mu  t}+B \mu  \,{e}^{\mu  t}}{A \,{e}^{2 \mu  t}+B \,{e}^{\mu  t}+C}-\frac{2 D \mu  \,{e}^{\mu  t}}{D \,{e}^{\mu  t}+1},
\end{equation}
where the corresponding coefficients are the same values obtained from the aforementioned fitting. Then, this function is compared with the Hubble parameter obtained from the direct solution of the Friedmann and the field equations. This comparison is shown in Fig. \ref{H2t2}, in which the solid line represents the numerical solution and the dotted line corresponds to the Hubble parameter obtained from the analytical parameterization. As can be seen from Fig. \ref{H2t2}, the two curves agree very well with each other at all epochs of the evolution of the universe, including the contraction phase, the transition region (bounce) and the expansion phase. In particular, both descriptions reproduce with reasonable accuracy the change of sign of the Hubble parameter at the bounce point, its maximum value at the beginning of the expansion phase, as well as its decreasing behavior at later times. The observed slight differences appear only in the regions with the most dynamic changes, which is due to the approximate nature of the parameterization function and does not have a noticeable effect on the overall behavior of the model.

Therefore, the purpose of introducing the Eq. \eqref{at2} is not to replace the field equations or to provide a new analytical solution for them, but rather to provide a precise analytical parameterization of the numerical solution that preserves the dynamical information of the model with high accuracy and allows its use in future analytical studies. Since the Hubble parameter is directly calculated from the scale factor and its time derivative, the very good agreement between the numerical results and the analytical parameterization for \(H(t)\) can be considered as an independent validation for the Eq. \eqref{at2}. Therefore, it is not necessary to make a similar comparison for other cosmological quantities, since these quantities are either obtained directly from the same background evolution and its derivatives or, as in the previous section, are calculated from the complete numerical solution of the field equations.

However, the description of the background evolution alone is not sufficient for a complete evaluation of the model, and its dynamical behavior and stability also need to be investigated. Therefore, in the next section, the structure of critical points, the stability of solutions, and the qualitative behavior of the evolution of the universe in phase space will be studied using an autonomous dynamical system.

%######################################################################
%######################################################################
\section{Dynamical System Analysis and Stability}\label{VI}

Here we explore the gravity model of non-minimal kinetic coupling by approach the dynamical system analysis. Therefore, we expect to find the stability conditions of the current system in a phase plane by using critical or fixed points. The standard form of the dynamical system is as $X'=f(X)$, where $X$ is a column vector for appropriate auxiliary variables and function $f$ is a column vector function related to the independent equations, and the prime symbol is the derivative with respect to $N = \ln a$. Therefore, critical points $X_c$ satisfy condition $X'=0$ and the linear stability properties of these critical points are determined.

For understanding of stability and instability fixed points, we put a system to an initial value that is close to its fixed point. If the trajectory of the solution of differential equation $X'=f(X)$ comes close to this fixed point, it is called a stable fixed point, and if it moves away from this fixed point, it is called an unstable fixed point. This means that a stable fixed point can be considered as an attractor and an unstable fixed point as a repeller. A particle controlled by $X'=f(X)$ forces the particle to move towards a stable fixed point, and an unstable fixed point will force a particle away from it.

To study the dynamics of the present system, we use the method of phase plane analysis, so we introduce the dimensionless variables as an autonomous dynamical system in the following form
\begin{subequations}\label{syst1}
\begin{eqnarray}
x=\frac{\kappa^2 \dot{\phi}^2}{6 H^2},~~~~~
y=\frac{\kappa^2 V}{3 H^2},~~~~~
z=6 \xi H^2 F, \label{syst1-1}\\
\chi=\frac{\kappa^2 \dot{V}}{ H^3},~~~~~
\psi=6 \xi H \dot{F},~~~~~ \Lambda=\frac{\kappa^2 \dot{\phi} \ddot{\phi}}{H^3},\label{syst1-2}
\end{eqnarray}
\end{subequations}

where the two main variables $x$ and $y$ related to kinetic energy and scalar field potential are chosen as independent variables of the phase space. The third variable $z$, which represents the non-minimal coupled term, can be defined as the dependent variable in terms of $x$ and $y$ using the following constraint equation \eqref{const1-1}, which is the first Friedman equation. The variables $\chi$, $\psi$ and $\Lambda$ are assumed to be constant in order to reduce the complications and to draw the phase space graphs more clearly and efficiently. These variables are model parameters that vary less than the two main variables $x$ and $y$ or their effect is applied in a parametric and constant manner. The assumption of the constancy of these variables allows the analysis of the phase space to be reduced to 2 dimensions and the computational complexity to be reduced, especially for the plotting of phase paths that must be fully representable and analyzable. In other words, the assumption that these parameters are constant means considering a ``cut'' or ``fixed cuts'' of the wider phase space so that the behavior of the main variables and the dominant dynamics can be better represented and more complex details can be controlled parametrically. Therefore, it is a common approach in the analysis of dynamical systems to consider some slow or infrequently changing parameters or variables as fixed parameters in order to maintain focus on the main dynamical variables and to make the stability analysis and phase space trajectories simple and understandable, where these attempt to analyze and scrutinize the model of the bouncing cosmology as an important feature. The important point that is evident here is that if the Hubble parameter becomes zero, what effect or changes would the autonomous variables have? Normally, the zero value for the Hubble parameter indicates the turning point in the early dynamics of the universe. Therefore, some autonomous variables, especially $x$ and $y$, tend to infinity and the progress of our calculations becomes undefined. To avoid this ambiguity, we should seek to limit these variables to a finite value, which in this case leads us to examine points after the neighborhood of the bounce point in the same early universe. For this reason, to advance this study, by replacing Eqs. \eqref{syst1} into Eqs. \eqref{fried2-1}, \eqref{fried2-3}, and \eqref{eosphi}, the following equations are obtained as
\begin{subequations}\label{const1}
\begin{eqnarray}
&x+y+3 x z = 1,\label{const1-1}\\
&\frac{\dot{H}}{H^2} = - \frac{\Lambda z +3 \psi  x+18 x z+\Lambda +\chi +18 x}{12 x z},\label{const1-2}\\
&\omega = -\frac{2 \Lambda  z +6 \psi  x +9 x z -9 x +9 y}{9 \left(3 x z +x +y \right)} - \frac{2 x z}{3 \left(3 x z +x +y \right)} \frac{\dot{H}}{H^2},\label{const1-3}
\end{eqnarray}
\end{subequations}
immediately obtain $\omega$ to insert \eqref{const1-1} and \eqref{const1-2} into \eqref{const1-3} as follows:
\begin{equation}\label{eosphi1}
\omega = 2 x - \frac{\psi  x}{2} - y + \frac{\Lambda  y}{18 x} - \frac{\Lambda}{18 x} + \frac{\Lambda}{9} + \frac{\chi}{18}.
\end{equation}

As we know, the EoS parameter is the relationship between energy density and pressure in a dominated fluid within the universe. This parameter helps us to better understand the different periods of the universe and is introduced as an observable quantity to describe the universe. These eras include the cosmic inflation era, the radiation dominance era, the matter dominance era, the structures formation era, and dark energy dominance era. In cosmic inflation period, the universe expanded so fast that the result was a homogeneous universe. A theory of modern particle physics states that particles in high density can surprisingly overcome gravity, that is, it causes gravity to be repulsive, contrary to its attraction, and this gravitational repulsion is powerful enough, which causes rapid expansion. As a result, the EoS parameter for this period is equal to $-1$. In the period of radiation dominance, which begins after the end of the inflation period, the energy density was dominated by radiation such as photons, neutrinos, and other relativistic particles, and its EoS parameter is $1/3$. After the radiation-dominant period, the matter-dominant period appears due to the expansion and cooling of the universe, in that case, the energy density is dominated by non-relativistic matter, so that its EoS parameter is equal to zero, which is usually also referred to as cold dark matter. In the structures formation period, with the continuation of the evolution of the universe, the gravitational attraction of matter caused the formation of the structures of the universe such as galaxies, galaxy clusters, and superclusters. In the period of dark energy dominance, which is related to the late universe, it has the EoS parameter smaller than $-1/3$, so that the observational data confirms the crossing of the cosmological constant value with an EoS equal to $-1$. We note that in this work, we focus on the early periods of the universe, especially the period of inflation.

To differentiate from Eqs. \eqref{syst1-1} with respect to $N = \ln{a}$, and to use from Eqs. \eqref{const1}, we earn the autonomous equations of the cosmic dynamic system as
\begin{subequations}\label{syst2}
\begin{eqnarray}
&\frac{dx}{d N}=\frac{-3 \psi x^2+\Lambda y-\chi x-12 x^2+6 x y-\Lambda-6 x}{2(x+y-1)}, \label{syst2-1}\\
&\frac{dy}{d N}=-\frac{9 \psi x^2 y+2 \Lambda x y-\Lambda y^2-2 \chi x^2+\chi x y+36 x^2 y-18 x y^2+\Lambda y+2\chi x+18 x y}{6 x (x+y-1)}.\label{syst2-2}
\end{eqnarray}
\end{subequations}

In order to describe the stability conditions, we have to solve the current autonomous system by setting $f(x, y)=dx/dN= 0$ and $g(x, y)=dy/dN= 0$ as the fixed points ${fp}_1$ and ${fp}_2$ in which $A=\sqrt{-3 \Lambda \psi-12 \Lambda+6 \chi+9}$ as shown in Tab. \ref{tab1}.

\begin{table}[h]
\caption{The fixed points (Critical points).} % title of Table
\centering % used for centering table
\begin{tabular}{||c |c |c ||} % centered columns (4 columns)
\hline\hline %inserts double horizontal lines
~$Points$~ & ${fp}_1$ & ${fp}_2$ \\ [0.5ex] % inserts table
%heading
\hline\hline % inserts single horizontal line
$x$ & \,\,\,\, $-\frac{\Lambda}{A+3}$\,\,\,\,\,\, & \,\,\,$\frac{\Lambda}{A-3}$~~~ \\
 \hline% inserting body of the table
$y$ &  $-\frac{\chi}{A+3}$ & $\frac{\chi}{A-3}$   \\
 [1ex] % [1ex] adds vertical space
\hline\hline %inserts single line
\end{tabular}
\label{tab1} % is used to refer this table in the text
\end{table}

Now to describe properties of the fixed points, we take linear perturbations for Eqs. \eqref{syst2} as $dx/dN \rightarrow dx/dN + \delta(dx/dN)$ and \(dy/dN \rightarrow dy/dN + \delta (dy/dN)\). Next, we should determine the eigenvalue of these fixed points (Tab. \ref{tab1}). For this purpose,  we write the system of differentials in the following matrix form
$$J=\begin{pmatrix}\label{matrix1}
\frac{\partial f}{\partial x} & \frac{\partial f}{\partial y} \\
\frac{\partial g}{\partial x} & \frac{\partial g}{\partial y}
\end{pmatrix}$$
where
\begin{subequations}\label{matrix2}
\begin{eqnarray}
&\frac{\partial f}{\partial x}=\frac{(3 \psi+12) x^{2} +(6 \psi+24)  x y +(\Lambda+\chi+12)  y -(6 \psi+24) x  -6 y^{2}-\Lambda -\chi -6}{-2 \left(-1+x +y \right)^{2}}, \label{matrix2-1}\\
&\frac{\partial f}{\partial y}=\frac{x \left(3 \psi  x +\Lambda +\chi +18 x \right)}{2 \left(-1+x +y \right)^{2}},\label{matrix2-2}\\
&\frac{\partial g}{\partial x}=\frac{-(9 \psi+54)  x^{2} y^2 +(2 \Lambda+3 \chi+9 \psi+54) x^{2} y-\Lambda y (2 x y+y^{2}-2 x-2 y+1)}{6 x^{2} \left(-1+x +y \right)^{2}},\label{matrix2-3}\\
&\frac{\partial g}{\partial y}=\frac{\left(9 \psi +36\right) x^{3}+\left(2 \Lambda +3 \chi -9 \psi -18\right) x^{2}-\left(36 x +2 \Lambda +18 y -36\right) x y -\left(\Lambda +3 \chi +18\right) x -\Lambda  \left(y^{2}-2 y +1\right)}{-6 x \left(-1+x +y \right)^{2}}.\label{matrix2-4}
\end{eqnarray}
\end{subequations}

Then, find the eigenvalues $\lambda_1$ and $\lambda_2$ by setting $det(J-\lambda I)=0$ as
\begin{subequations}\label{eigen1}
\begin{eqnarray}
&\lambda_1=-\frac{9 x^{2} \psi +2 \Lambda  x -\Lambda  y +3 \chi  x +36 x^{2}-18 x y +\Lambda +18 x}{6 x \left(-1+x +y \right)}, \label{eigen1-1}\\
&\lambda_2=\frac{-3 x^{2} \psi -3 \psi  x y +6 \psi  x -12 x^{2}-6 x y +6 y^{2}+\Lambda +\chi +24 x -12 y +6}{2 \left(-1+x +y \right)^{2}},\label{eigen1-2}
\end{eqnarray}
\end{subequations}
where related to parameters $x$, $y$, $\chi$, $\psi$, and $\Lambda$. To substitute the amount $x$ and $y$ fixed points from Tab. \ref{tab1}, the eigenvalues obtain as shown in Tab. \ref{tab2}. We note that the our dynamical system analysis depends on the eigenvalues, which are studied as follows:
\begin{itemize}
\item For real and positive eigenvalues ($\lambda_1 > 0$ and $\lambda_2 > 0$), trajectories move away from the critical point or fixed point, i.e., the track has a source, one's introduced as an unstable node.
\item For real and negative eigenvalues ($\lambda_1 < 0$ and $\lambda_2 < 0$), trajectories move towards the critical point or fixed point, i.e., the track has a sink, one's introduced as an stable node.
\item For real and opposite eigenvalues ($\lambda_1 > 0$ and $\lambda_2 < 0$ and vice versa), trajectories move both inwards and outward the critical point or fixed point, i.e., the trajectory simultaneously has a sink and has a source, and is introduced as the saddle point.
\end{itemize}

\begin{table}[h]
\caption{Stability conditions of critical points.} \label{tab2}
\begin{NiceTabular}{||c|c|c|c||}
  \hline\hline
  \rm{Fixed points} & \rm{Eigenvalues} & \rm{Acceptable conditions} & \rm{Stability conditions}\\
  \hline\hline
                    &  &  & \\
                    & $\lambda_1=\frac{A}{3}+1$ & $A < -3$ ~(\rm{impossible}) &\\
                    &  &  &\\
                   \cline{2-3}
  {${fp}_1$}&  &  & ~~\rm{Unstable}~~\\
                    & ~~$\lambda_2=\frac{A (A+3)}{\Lambda+\chi+A+3}$~~ & ~~$A>0,~\Lambda+\chi < -(A+3)$~ & \\
                   &  &  &\\
  \hline \hline
                    &  &  &\\
                    & $\lambda_1=-\frac{A}{3}+1$ & $A>3$ &\\
                    &  & & \\
                     \cline{2-3}
  {${fp}_2$}&  &  & \rm{Stable}/\rm{Unstable}\\
                    & $\lambda_2=\frac{A (A-3)}{\Lambda+\chi-A+3}$ & $A>3,~ \Lambda+\chi < A-3$ & \\
                    &  & $ 0<A<3,~ \Lambda+\chi > A-3$ & \\
                    &  &  &\\
\hline\hline
\end{NiceTabular}
\end{table}

According to the above, we obtain acceptable conditions for eigenvalues as shown in Tab. \ref{tab2}. We note that $A$ is always positive, so we have $-\Lambda \psi-4 \Lambda+2 \chi+3 > 0$, it means, the obtained condition is the constraint between the parameters $\chi$, $\psi$, and $\Lambda$. However, by substituting the fixed points in Tab. \ref{tab1} into Eqs. \eqref{eigen1}, the results in the second column of Tab. \ref{tab2} are calculated. As a result, we find acceptable conditions for fixed points $fp_1$ and $fp_2$ based on negative eigenvalues and insert the results in the third column of Tab. \ref{tab2}. The acceptable condition $fp_1$ exists only for eigenvalue $\lambda_2$ as $A>0$ and $\Lambda+\chi<-(A+3)$. The acceptable condition $fp_2$ exists for the eigenvalue $\lambda_1$ is only one condition in the form $A>0$, and for the eigenvalue $\lambda_2$ there are two conditions in the forms $A>3, \Lambda+\chi<A-3$ and $0<A<3, \Lambda+\chi>A-3$. We should note that in this work, we study the early universe and focus on the inflationary era immediately after the bounce.

\begin{figure}[hh]
\begin{center}
\includegraphics[scale=.45]{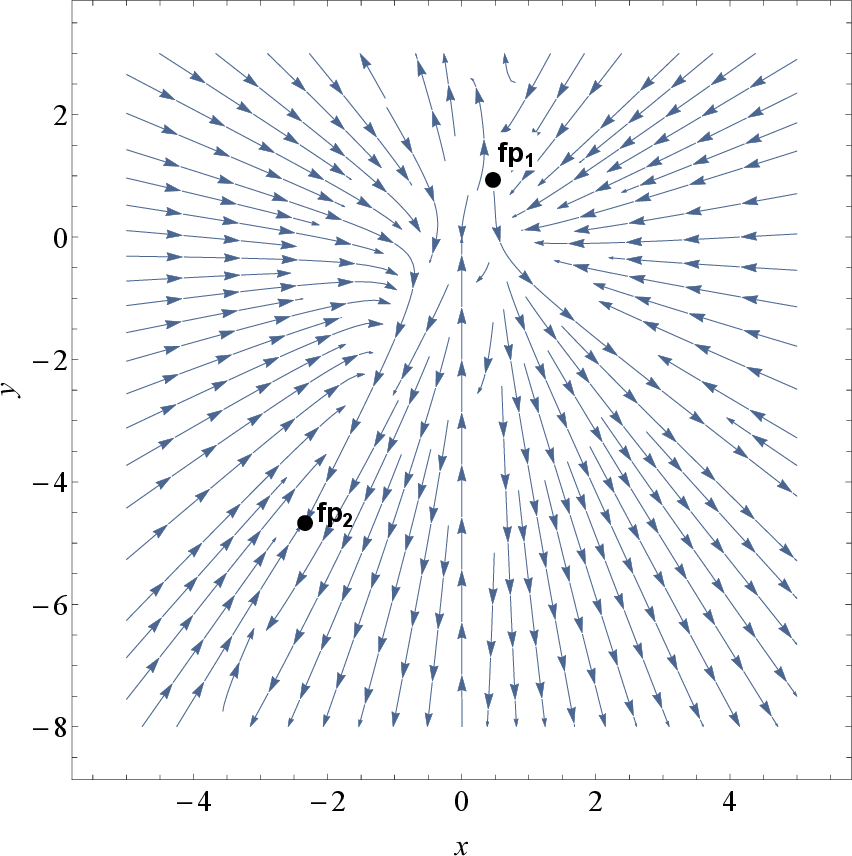}~~~~~~\includegraphics[scale=.45]{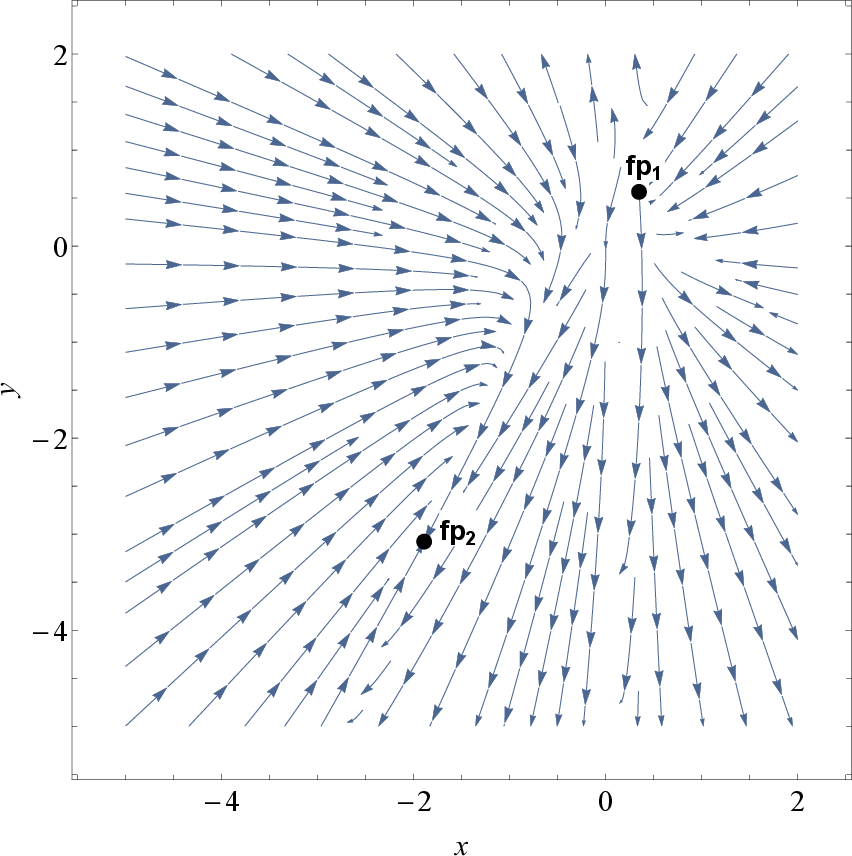}
\caption{Left panel: Phase space trajectories on the $x-y$ plane for the values $\Lambda=-3.5$, $\chi=-7$ and $\psi=1.07$. Right panel: Phase space trajectories on the $x-y$ plane for the values $\Lambda=-2.55$, $\chi=-4.15$ and $\psi=0.55$.}\label{fp12}
\end{center}
\end{figure}

To analyze the behavior and stability of a dynamic system, we use a phase portrait, which is a valuable tool in the study of dynamic systems. The trajectories of a dynamic system in the phase plane are represented geometrically by the phase portrait. A curve or point is drawn with the initial conditions defined for each set. To analyze the phase paths in $x-y$ plane, the parameters $\chi$, $\psi$, and $\Lambda$ were considered as structural parameters of the fixed model. This choice was made in order to reduce the dimensions of the dynamic space and to allow for a qualitative examination of the system behavior. In each diagram below, the effect of changing these parameters is examined separately. For this purpose, we find the stability conditions of the inflation era according to the fourth column of Tab. \ref{tab2}, which we describe how to obtain it separately for different modes as follows:

\textbf{The first mode:}
Since $A$ is always positive, so $\lambda_1$ is also always positive, and $\lambda_2$ will be negative for the acceptable condition, i.e, $A>0$ and $\Lambda+\chi<-(A+3)$. In order to represent attractor and repeller in the phase space,  based on the aforesaid acceptable condition, we choose the values of $\Lambda=-3.5$, $\chi=-7$, and $\psi=1.07$, and plot phase portrait of the equations system \eqref{syst2} in $x-y$ plane according to the left panel of the Fig. \ref{fp12}. From this figure, we find that $fp_1$ is a saddle point or repeller, but in contrast, $fp_2$ is a stable point or attractor. We note that the values of EoS obtain by Eq. \eqref{eosphi1} for the fixed points as $\omega(fp_1) \simeq -1$ and $\omega(fp_2) \simeq 0$.

\textbf{The second mode:}
In this mode, we use the acceptable condition of $A>3$. By this condition, we choose the values of $\Lambda=-2.55$, $\chi=-4.15$, and $\psi=0.55$, and plot the equations system \eqref{syst2} according to the right panel of Fig. \ref{fp12}. The figure shows that $fp_1$ is a saddle point or repeller, but, $fp_2$ is a stable point or attractor. Also, We can obtain the values of EoS for specified fixed points as $\omega(fp_1) \simeq -0.3$ and $\omega(fp_2) \simeq -1$.

\begin{figure}[hh]
\begin{center}
\includegraphics[scale=.45]{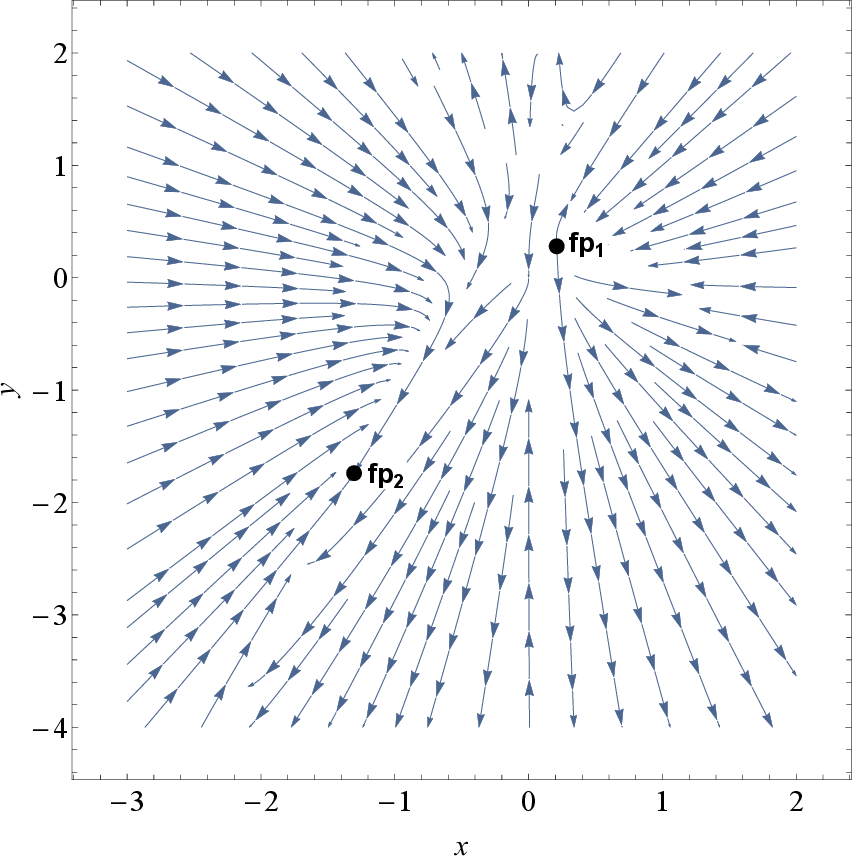}~~~~~~\includegraphics[scale=.45]{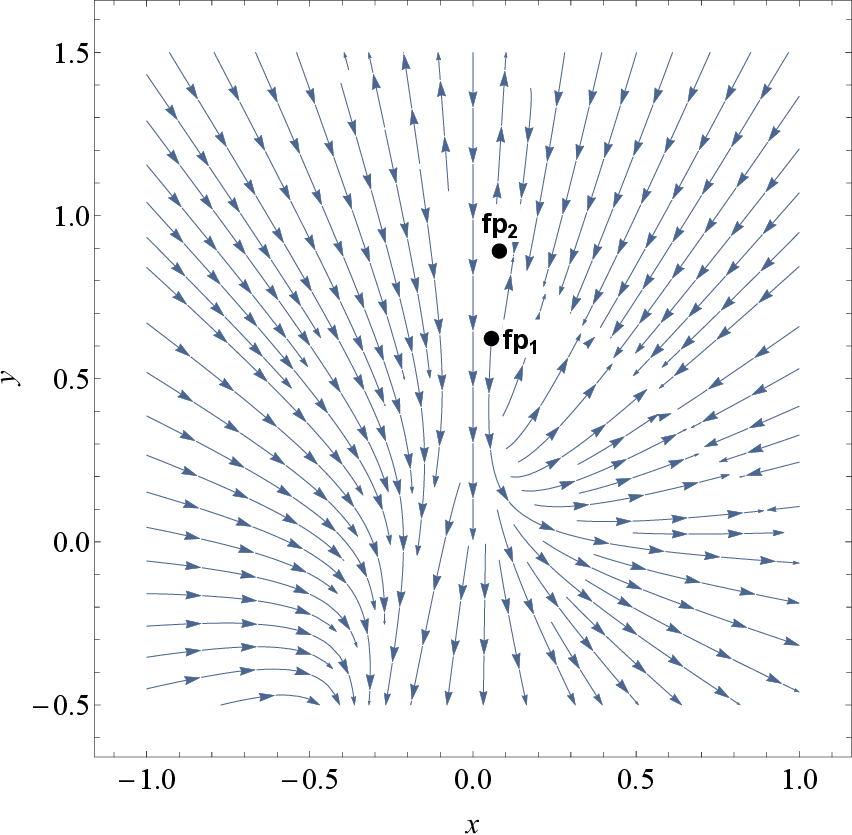}
\caption{Left panel: Phase space trajectories on the $x-y$ plane for the values $\Lambda=-1.5$, $\chi=-2$ and $\psi=0.494$. Right panel: Phase space trajectories on the $x-y$ plane for the values $\Lambda=-0.2$, $\chi=-2.2$ and $\psi=3.47$.}\label{fp212}
\end{center}
\end{figure}

\textbf{The third mode:}
Herein, we consider the acceptable condition of $A>3$ and $\Lambda+\chi<A-3$ from Tab. \ref{tab2}. By applying this condition, we choose the values of $\Lambda=-1.5$, $\chi=-2$, and $\psi=0.494$, and plot the equations system \eqref{syst2} according to the left panel of Fig. \ref{fp212}. From this figure we find that $fp_1$ is a saddle point or repeller, and $fp_2$ is a stable point or attractor. In addition, the values of EoS obtain for the fixed points as $\omega(fp_1) \simeq 0.1$ and $\omega(fp_2) \simeq -1$.

\textbf{The fourth mode:}
In this mode, considering the acceptable condition $0<A<3$ and $\Lambda+\chi>A-3$ in Tab. \ref{tab2}. it leads us to choose the values $\Lambda=-0.2$, $\chi=-2.2$, and $\psi=3.47$, and plots the equations system \eqref{syst2} according to the right panel of Fig. \ref{fp212}. As can be seen, Fig. \ref{fp212} shows that both fixed points $fp_1$ and $fp_2$ are saddles or repellers. So that the EoS values for fixed points are obtained as $\omega(fp_1) \simeq -0.68$ and $\omega(fp_2) \simeq -1$.

In general, the aforesaid modes show an overview of the early universe, this means that after bounce point, the universe enters different periods. One of these periods is cosmic inflation with EoS equal to $-1$. For this purpose, the motivation of our choice for the values $\Lambda$, $\chi$, and $\psi$ is to establish the acceptable conditions (containing Tab. \ref{tab2}) and $\omega=-1$. In fact, the shape of the inflation potential has a direct effect on the stability of cosmic inflation. For this reason, we choose the potential \eqref{pot1} to satisfy the above description that even the phase space trajectories show the stability of the universe in the inflationary period. Therefore, the obtained results bring us a more accurate understanding of the early period of the universe.

Therefore, fixed points in phase space represent states in which the rate of change of dynamical variables becomes zero. These states can represent different phases of the universe, such as:
\begin{itemize}
\item A phase of accelerated expansion (e.g. with $\omega \approx -1$) which is similar to the inflationary or dark energy era.
\item A phase of slow or fast contraction (e.g. with $\omega > 0$) which can be a pre-bounce.
\item A critical bouncing point where the Hubble parameter $H=0$ and $\dot{H}>0$, represents the transition from contraction to expansion.
\end{itemize}
For each fixed point, in addition to the EoS value, we have investigated the behavior of physical quantities such as the Hubble parameter $H$, the scalar field $\phi$, its derivative $\dot{\phi}$, the energy density, and the effective pressure. Also, by analyzing the Jacobian matrix around the fixed points, their type of stability (attractive, repulsive, saddle) is determined, and the phase trajectories in Figs. \ref{fp12} and \ref{fp212} show how the system approaches or passes through these points. As a result, the fixed points represent key stages in the evolution of the universe not only mathematically, but also physically including the beginning of expansion, the end of contraction, and the bouncing transition.

%%%%%%%%%%%%%%%%%%%%%%%%%%%%%
%%%%%%%%%%%%%%%%%%%%%%%%%%%%%

\section{Conclusion}\label{VII}

In this paper, the evolution of the universe in the framework of the bouncing cosmology was studied using the non-minimal kinetic coupling theory. For this purpose, we started from a flat FRW universe and by applying the condition \(\eta+2\xi=0\), the model action was reduced to the derivative coupling form with the Einstein tensor. Then, the modified Friedmann equations and the field equation were extracted and based on them, the corresponding relations for the energy density and effective pressure were obtained. Also, by choosing a suitable potential for the scalar field and a polynomial coupling function, the necessary framework for studying the dynamics of the bouncing universe was provided.

In what follows, by applying the Bounce condition, the system of Friedmann equations and the field equation was solved numerically simultaneously and without assuming any analytical form for the scale factor. This numerical solution determined the time evolution of the scale factor, the scalar field and other background quantities directly from the dynamic equations of the model. Then, in order to provide an explicit analytical representation of this background evolution, an analytical parameterization for the scale factor was introduced, which was fitted to the numerical solution with very high accuracy. The validity of this parameterization was also assessed by comparing the resulting Hubble parameter with the Hubble parameter obtained from the numerical solution, and the very good agreement between the two indicated that the proposed function can be used as a reliable analytical approximation of the background evolution in this model. Although this parameterization is not derived from the field equations, it preserves the dynamical information of the numerical solution with high accuracy and can be used in further studies, including the investigation of cosmological perturbations, the calculation of observational quantities, thermodynamic analyses, and other analytical applications. In addition, numerical investigation of the Hubble parameter, the comoving Hubble horizon, the energy density, the pressure, and the EoS parameter showed that the present model is able to describe the smooth transition from the contraction phase to the expansion phase without the occurrence of singularities and reproduces the expected qualitative behavior of a bouncing universe.

To explore the stability of the model, we used dynamical system analysis in phase plane by critical or fixed points. For the same purpose, we introduced the dimensionless variables as an autonomous dynamical system, and then we earned the autonomous equations of the cosmic dynamic system. Afterward, by setting the autonomous equations equal to zero, we obtained the fixed points according to Tab. \ref{tab1}, next, stability conditions of critical points were obtained according to the contents of Tab. \ref{tab2}. In what follows, we considered four different modes for the fixed points and then, obtained the stability points by placing constraints for each mode. After that, we drew the figures of the phase space trajectories in $x-y$ phase plane. The results obtained from these modes showed that, after the bounce point, the universe enters different periods, one of which is cosmic inflation. Since the shape of the potential has a direct effect on the stability in the inflationary period, so, choosing potential \eqref{pot1} brought good results for the inflationary period. Finally, the results obtained help us to better understand the early universe and even the evolution of the universe as a viable model.

Finally, although the analytical parameterization presented in this paper was able to reproduce the background evolution obtained from the numerical solution with high accuracy, examining the relationship between the coefficients of this parameterization and the fundamental parameters of the model can lead to a deeper understanding of their physical origin. Therefore, extracting these dependencies and providing a physical interpretation of the parameterization coefficients will be one of the valuable paths for the development of future research. In addition, having an explicit analytical form for the scale factor can provide wide applications, including the investigation of scalar and tensor perturbations, the study of early gravitational waves, the reconstruction of modified gravity models, and the thermodynamic analysis of bouncing universes. Overall, the framework presented in this paper is expected to be used as a suitable platform for developing semi-analytical methods in complex cosmological models.

%%%%%%%%%%%%%%%%%%%%%%%%%%%%%
%%%%%%%%%%%%%%%%%%%%%%%%%%%%%

\end{document}